\documentclass[fleqn,preprint]{elsarticle}
\AtBeginDocument{\setlength{\mathindent}{1pt}}

\usepackage{booktabs}
\usepackage{longtable}
\usepackage[flushleft]{threeparttable}
\usepackage[flushleft]{threeparttablex}
\usepackage{multirow}
\usepackage{multicol}
\usepackage{amsmath,amssymb,amsfonts}
\usepackage{algorithm}
\usepackage{array}
\usepackage{textcomp}
\usepackage{stfloats}
\usepackage{url}
\usepackage{verbatim}

\usepackage{eurosym}
\usepackage{acronym}
\usepackage[mathscr]{eucal}
\usepackage{xcolor}
\usepackage{subfigure}
\usepackage{enumitem}
\usepackage{accents}
\usepackage{amssymb}
\usepackage{natbib}
\usepackage{geometry}
\usepackage{graphicx}
\usepackage{txfonts}
\usepackage{hyperref}
\usepackage{float}
\usepackage{tcolorbox}
\tcbuselibrary{skins,breakable}

\biboptions{sort&compress}

\usepackage{subcaption} % For subfigure support
\usepackage{caption}    % For better control over captions
\usepackage{url}
\usepackage{breakurl}

\usepackage{makecell}   % Allows text wrapping in table cells

\usepackage{pifont} % Include in the preamble

\usepackage{rotating}
\usepackage[nameinlink]{cleveref}

\usepackage{tabularx}

\usepackage{tikz}
\usetikzlibrary{shapes.geometric, arrows.meta, positioning}
\tikzstyle{block} = [rectangle, draw, text centered, minimum height=1.2em, font=\small, rounded corners, fill=blue!10]
\tikzstyle{decision} = [diamond, draw, text centered, inner sep=0pt, minimum width=2cm, font=\small]
\tikzstyle{input} = [rectangle, draw, text centered, font=\small, fill=gray!20]
\tikzstyle{uncertain} = [ellipse, draw, dashed, fill=red!10, font=\small]
\tikzstyle{arrow} = [thick,->,>=stealth]

\usepackage{tikz}
\tikzset{
  panel label/.style={
    anchor=north west,
    font=\small,
    fill=white, fill opacity=0.75, text opacity=1,
    inner sep=2pt,
    xshift=3mm, yshift=-3mm
  }
}

\usepackage{algpseudocode}

\usetikzlibrary{arrows.meta,positioning,shapes.geometric,fit,calc}

\newcommand{\ubar}[1]{\underaccent{\bar}{#1}}

\usepackage{nomencl}

\makenomenclature

\usepackage{xstring}
\usepackage{etoolbox}
\renewcommand\nomgroup[1]{%
  \item[\bfseries
  \ifstrequal{#1}{I}{Indexes and Sets}{%
  \ifstrequal{#1}{P}{Parameters}{%
  \ifstrequal{#1}{V}{\textcolor{black}{Variables}}{%
  \ifstrequal{#1}{W}{Binary Variables}{%
  \ifstrequal{#1}{Z}{Vectors}{}}}}}%
]}

\graphicspath{{figs/}} 

\acrodef{sb}[SB]{Single-Bound}
\acrodef{mb}[MB]{Multi-Bound}
\acrodef{mbro}[MBRO]{Multi-Bound Robust Optimization}
\acrodef{dam}[DAM]{Day-Ahead Market}
\acrodef{srm}[SRM]{Secondary Reserve Market}
\acrodef{ess}[ESS]{Energy Storage System}
\acrodef{res}[RES]{Renewable Energy Sources}
\acrodef{der}[DER]{Distributed Energy Resources}
\acrodef{ndrs}[ND-RES]{Non-dispatchable RES}
\acrodef{drs}[D-RES]{Dispatchable RES}
\acrodef{vpp}[VPP]{Virtual Power Plant}
\acrodef{pv}[PV]{Photovoltaic}
\acrodef{milp}[MILP]{Mixed Integer Linear Programming}
\acrodef{rVPP}[RVPP]{Renewable-based VPP}
\acrodef{rvpp}[RVPP]{Renewable-based Virtual Power Plant}
\acrodef{gams}[GAMS]{General Algebraic Modeling System}
\acrodef{so}[SO]{Stochastic Optimization}
\acrodef{ro}[RO]{Robust Optimization}
\acrodef{aro}[ARO]{Adaptive Robust Optimization}
\acrodef{saro}[SARO]{Stochastic Adaptive RO}
\acrodef{dro}[DRO]{Distributed RO}
\acrodef{ccg}[C\&CG]{Column \& Constraint Generation}
\acrodef{ev}[EV]{Electric Vehicle}
\acrodef{igdt}[IGDT]{Information Gap Decision Theory}
\acrodef{cvar}[CVaR]{Conditional Value-at-Risk}
\acrodef{vab}[VaB]{Value-at-Best}
\acrodef{soc}[SoC]{State of Charge}
\acrodef{pvt}[PVT]{Photovoltaic-Thermal}
\acrodef{woa}[WOA]{Whale Optimization Algorithm}
\acrodef{dr}[DR]{Demand Response}
\acrodef{chp}[CHP]{Combined Heat and Power}
\acrodef{cco}[CCO]{Chance-Constrained Optimization}
\acrodef{rec}[REC]{Renewable Energy Certificate}
\acrodef{cer}[CER]{Carbon Emission Right}
\acrodef{mc}[MC]{Marginal Contributions}

\journal{Applied energy}

\ExplSyntaxOn
\NewDocumentCommand{\instringTF}{mmmm}
{\oleks_instring:nnnn { #1 } { #2 } { #3 } { #4 }}
\tl_new:N \l__oleks_instring_test_tl
\cs_new_protected:Nn \oleks_instring:nnnn
{
    \tl_set:Nn \l__oleks_instring_test_tl { #1 }
    \regex_match:nnTF { \u{l__oleks_instring_test_tl} } { #2 } { #3 } { #4 }
}
\ExplSyntaxOff
\let\mybibitem\bibitem
\renewcommand{\bibitem}[1]{%
    \instringTF{Refblack}{#1}
    {\color{black}\mybibitem{#1}}
    {\color{black}\mybibitem{#1}}%
}

\begin{document}

\begin{frontmatter}

\title{A Multi-Bound Robust Optimization Approach for Renewable-Based VPP Market Participation Considering Intra-Hourly Uncertainty Exposure}

%Multi-Bound Robust Bidding for Renewable VPPs in Intra-Hourly Electricity Markets

%A Multi-Bound Robust Optimization Model for Renewable-Based VPP Participation in Transitioning Hourly to Quarter-Hourly Electricity Markets

%A Multi-Bound Robust Bidding Strategy for Renewable-Based VPPs Considering Intra-Hourly Uncertainty Exposure

\author{Hadi Nemati*}

%\affiliation{organization={Comillas Pontifical University ICAI School of Engineering, Institute for Research in Technology},%Department and Organization
%            state={Madrid},
%            country={Spain}} 

\author{{\'A}lvaro Ortega}
\author{Enrique Lobato}
\author{Luis Rouco}

\address{Comillas Pontifical University, Institute for Research in Technology, Madrid, Spain}    

\address{*Corresponding author}
\address {E-mail address: hnemati@comillas.edu}

\begin{abstract}

{\color{black}With the ongoing transition of electricity markets worldwide from hourly to intra-hourly bidding, market participants—especially \ac{res}—gain improved opportunities to adjust energy and reserve schedules and to benefit from more accurate higher-resolution forecasts.} However, this shift requires participants to update decision-making frameworks and to strengthen uncertainty management in order to fully exploit the new market potential. In particular, Renewable-Based Virtual Power Plants (RVPPs) aggregating dispatchable and non-dispatchable RES must account for these changes through market-oriented scheduling methods that efficiently address multiple uncertainties, including electricity prices, RES generation, and demand consumption. In this vein, this paper proposes a multi-bound robust optimization framework to simultaneously capture these uncertainties, explicitly incorporate intra-hourly variability, and differentiate the deviation levels (frequent, moderate deviations and rare, extreme ones) of uncertain parameters. The proposed approach yields less conservative and more implementable bidding and scheduling decisions, thus improving RVPP profitability in both energy and reserve markets. Simulation studies compare the proposed method with standard robust optimization and evaluate the operational, market-strategy, and economic impacts of quarter-hourly versus hourly market resolution. Results indicate that the normalized absolute differences, across different uncertainty-handling strategies, between hourly and 15-minute schedules are 18.0–34.2\% for day-ahead traded energy, and 28.7–65.6\% and 10.1–16.3\% for upward and downward reserve traded in the secondary reserve market, respectively. Furthermore, relative to classic robust optimization, the proposed multi-bound approach increases profit by 24.9–49.2\% across the considered strategies.

\end{abstract}

\end{frontmatter}

%\vspace{.4cm}
\section*{Keywords}
Renewable-based virtual power plant, Multi-bound robust optimization, Intra-hourly uncertainty, energy and reserve markets

\vspace{2cm}
%\section*{Nomenclature}
%This subsection presents the notation and nomenclature used in the remainder of the paper.

\begin{tcolorbox}[breakable,colback=white, colframe=black,title=Nomenclature, width=\textwidth]
    
\begin{footnotesize}

\noindent \textbf{Abbreviations}

\begin{tabular}{p{1.1cm} p{4.8cm} p{1.1cm} p{5.2cm}}
\textbf{} & \textbf{} & \textbf{} & \textbf{} \\ 
C\&CG  & Column \& Constraint Generation            & ND-RES & Non-Dispatchable Renewable Energy Sources \\
CVaR   & Conditional Value-at-Risk                  & PV     & Photovoltaic \\
D-RES  & Dispatchable Renewable Energy Sources      & RES    & Renewable Energy Sources \\
DAM    & Day-Ahead Market                           & RO     & Robust Optimization \\
DRO    & Distributed Robust Optimization            & RVPP   & Renewable-Based Virtual Power Plant \\
ESS    & Energy Storage System                      & SO     & Stochastic Optimization \\
EV     & Electric Vehicle                           & SRM    & Secondary Reserve Market \\
IGDT   & Information Gap Decision Theory            & VPP    & Virtual Power Plant \\
MILP   & Mixed Integer Linear Programming           &        &  \\
\end{tabular}
\vspace{1em}

\noindent \textbf{Indices and Sets}
\vspace{.5em}

\renewcommand{\arraystretch}{1.15}
\begin{tabularx}{\textwidth}{@{}lX@{}}
$t \in \mathscr{T}$ \hspace{3.5mm} & Set of time periods \\
$u \in \mathscr{U}$ & Set of \ac{rvpp} units, comprising \acp{ndrs} ($r \in \mathscr{R}$), \acp{ess} ($s \in \mathscr{S}$), \acp{drs} ($c \in \mathscr{C}$), and demands ($d \in \mathscr{D}$) \\
$k \in \mathscr{K}$ & Set of uncertainty bounds in \ac{mbro} approach \\
$t \in \mathscr{T}_k$ & Set of worst-case time periods for bound $k$ \\
$\Xi^{F/S}$ & Sets of first- and second-level decision variables of the \ac{rvpp} operator \\

\end{tabularx}
\vspace{.5em}

\noindent \textbf{Parameters}
\vspace{.5em}

\begin{tabularx}{\textwidth}{@{}lX@{}}
$C_u$ \hspace{6mm} & Marginal operation cost of unit $u$ \hfill [€/MWh] \\
$E_u$ & Rated energy capacity of unit $u$ \hfill [MWh] \\
$P_u$ & Rated power capacity or power forecast of unit $u$ \hfill [MW] \\
$R_{u}$ & Ramp rate of unit $u$ \hfill [MW/min] \\
$T^{SR}$ & Secondary reserve activation time \hfill [min] \\
$\beta_u$ & Reserve provision limit factor of unit $u$ \hfill [\%] \\
$\Gamma_k$ & Uncertainty budget for bound $k$ \hfill [--] \\
$\Delta t$ & Time step duration \hfill [h] \\
$\eta_u$ & Efficiency of unit $u$ \hfill [\%] \\
$\lambda_t$ & Forecast electricity market price during period $t$ \hfill [€/MW, €/MWh] \\
\end{tabularx}
\vspace{.5em}

\noindent \textbf{Variables}
\vspace{.5em}

\begin{tabularx}{\textwidth}{@{}lX@{}}
$e_{u,t}$ \hspace{6mm} & Stored energy level of unit $u$ at the end of period $t$ \hfill [MWh] \\
$p_t^{DA}$ & Power traded by the \ac{rvpp} in the \acs{dam} during period $t$ \hfill [MW] \\
$p_{u,t}$ & Power dispatch of unit $u$ during period $t$ \hfill [MW] \\
$r_t^{SR}$ & Reserve capacity traded by the \ac{rvpp} in the \ac{srm} during period $t$ \hfill [MW] \\
$r_{u,t}$ & Reserve contribution provided by unit $u$ during period $t$ \hfill [MW] \\
$\sigma_{u}$ & Fraction of unit $u$ capacity allocated for reserve \hfill [\%] \\
$y_{k,t}$ & Auxiliary variable for uncertainty bound $k$ during period $t$ \hfill [MW, MWh] \\
$z_{k,t}$ & Auxiliary variable for uncertainty budget of bound $k$ during period $t$ \hfill [--] \\
$\phi_k$ & Dual variable associated with the uncertain parameters of bound $k$ \hfill [€, MW] \\
$\zeta_t$ & Dual variable associated with uncertain parameters during period $t$ \hfill [€, MW] \\
$v_{u,t}$ & Binary commitment status of unit $u$ during period $t$ \hfill [--] \\
$\chi_{u,k,t}$ & Binary indicator for worst-case realization of unit $u$ under bound $k$ during period $t$ \hfill [--] \\
\end{tabularx}
%\vspace{3em}

\vspace{8em}
\noindent \textbf{Superscripts and Accents}
%\vspace{.5em}

\begin{tabularx}{\textwidth}{@{}lX@{}}
$A^{DA}$ \hspace{4.7mm} & Related to the \ac{dam} \\
$A^{SR}$ & Related to the \ac{srm} \\
$A^\uparrow, A^\downarrow$ & Upward and downward reserve directions \\
$A^+, A^-$ & Charging and discharging states of \acp{ess} \\
%$\bar{A}, \ubar{A}$ & Upper and lower limits \\
$\tilde{A}$ & Median forecast value of an uncertain parameter \\
$A^{*}$ & Optimal value of a variable \\
$\hat{A}, \check{A}$ & Positive and negative deviations of an uncertain parameter from the forecast \\
${\bar{A}}, {\ubar{A}}$ & Upper and lower bounds (e.g., limits or forecast bounds) of a parameter \\
\end{tabularx}
\vspace{-1em}
\end{footnotesize}
\end{tcolorbox}

\acresetall %resets all acronym definitions
\section{Introduction}
\subsection{Motivation}

{\color{black}Electricity markets worldwide are currently undergoing a fundamental structural transition toward higher temporal granularity, moving from traditional hourly settlements to sub-hourly (e.g., quarter-hourly) intervals. This shift is driven by evolving market rules aiming to harmonize market designs and, more critically, to facilitate the seamless integration of variable \ac{res}~\cite{prete2025time}.} Within Europe, one recent example is the Spanish market's scheduled transition in October 2025, which replaces the standard 24 daily price points with 96 quarter-hourly intervals~\cite{EC_Energy_Market, REE2025}. The primary goal of this transition is to align market schedules more closely with physical grid operations, incentivizing this way participants to balance supply and demand in near real-time and reducing reliance on costly adjustment mechanisms.

Under this transformed regulatory framework, the core platform for energy trading, the \ac{dam}, evolves from the traditional 24-period structure to a higher resolution timeline divided into 96 quarter-hourly intervals, in which generation and demand agents submit bids for each 15-minute slot, after which the market operator establishes market-clearing prices that reflect the immediate volatility of the grid more accurately than hourly averages~\cite{omie_homepage}. Parallel to energy trading, the \ac{srm}, particularly for automatic frequency restoration reserve, remains the critical mechanism for the system operator to restore system frequency and power exchanges to scheduled values~\cite{entsoe_balancing}. However, the technical requirements for providing reserve capacity are now tightly synchronized with the quarter-hourly energy settlement periods.

%To participate effectively, the \ac{vpp} must leverage the fast-response capabilities of its internal assets—such as \ac{ess} and dispatchable units—to provide continuous reserve capacity within these shorter windows, maximizing profitability by co-optimizing energy and reserve commitments across 96 daily decision points~\cite{ruan2024data}.

{\color{black}A \ac{rvpp} operates as a unified entity that aggregates diverse assets, ranging from stochastic \ac{ndrs} and controllable \ac{drs} to \ac{ess} and flexible demand, to provide grid services and enhance profitability~\cite{yang2023optimal}. For market participants, particularly \acp{rvpp}, this transition brings both opportunity and challenge. The shift to finer time resolutions exposes \acp{rvpp} to "intra-hourly" dynamics previously smoothed out in hourly models. While market structures have evolved to accommodate \ac{res} variability, traditional optimization models often lag behind, relying on coarse uncertainty handling approaches that miss rapid fluctuations in \ac{res} production, demand, and \ac{dam}/\ac{srm} prices at a 15-minute scale~\cite{gao2024review}. Consequently, fully leveraging high-resolution markets requires updating \ac{rvpp} bidding strategies~\cite{kaiss2025review}. Specifically, uncertainty handling frameworks must be refined beyond hourly assumptions, ensuring that the \ac{rvpp} can manage and profit from the granular volatility inherent in this new market era.}

\subsection{Literature Review}

The participation of \acp{vpp} in the \ac{dam} has been extensively researched, particularly regarding the management of uncertainties associated with \ac{res} production and demand. Numerous studies employ \ac{ro} and \ac{so} techniques to address these challenges~\cite{liu2025two, wu2024two, feng2024optimal, li2023robust, falabretti2023scheduling, cao2025scenario}. For instance, Liu et al.~\cite{liu2025two} proposed a two-stage \ac{ro} approach that incorporates a refined demand response strategy, solving the problem via the \ac{ccg} algorithm with fuzzy-based subproblems. Wu et al.~\cite{wu2024two} developed a two-stage \ac{dro} model for rural \acp{vpp} integrating biomass and \ac{res}, utilizing a dual vertices fixing algorithm to handle correlated wind and solar uncertainties. The integration of \ac{ess} was addressed by Feng et al.~\cite{feng2024optimal}, who applied an improved nested \ac{ccg} algorithm within a two-stage \ac{dro} framework to manage power deviations. Beyond electricity, Li et al.~\cite{li2023robust} extended \ac{ro} to multi-energy \acp{vpp} participating in both energy and peak-regulation markets. Falabretti et al.~\cite{falabretti2023scheduling} utilized a Monte Carlo-based two-stage \ac{so} for \acp{rvpp} with \ac{ev} charging stations, while Cao et al.~\cite{cao2025scenario} explored carbon trading flexibility using scenario-based \ac{dro}. However, a significant limitation of these works is the neglect of market price uncertainty, a factor that critically impacts \ac{vpp} profitability.

To address this limitation, recent literature has incorporated market price uncertainty alongside generation variability~\cite{wang2024optimal, mei2024two, xiao2024windfall, shafiekhani2022optimal, afzali2025flexibility}. Wang et al.~\cite{wang2024optimal} introduced a two-stage \ac{dro} model that handles \ac{dam} price uncertainty in the first stage and utilizes \ac{cvar} to manage real-time price risks. Incorporating risk preferences, Xiao et al.~\cite{xiao2024windfall} proposed an \ac{so} framework capable of modeling both risk-seeking and risk-averse strategies, while Shafiekhani et al.~\cite{shafiekhani2022optimal} employed \ac{igdt} to balance economic viability and emission minimization under different risk attitudes. Advanced bi-level structures have also been proposed; Afzali et al.~\cite{afzali2025flexibility} developed a risk-aware \ac{so}--\ac{ro} strategy for energy and flexibility markets, reformulating the problem as a \ac{milp} to ensure global optimality. Furthermore, hybrid uncertainty handling methods have emerged, such as the data-driven \ac{ro} model by Ma et al.~\cite{ma2025data}, which captures price uncertainty via data-driven techniques and physical uncertainties via \ac{ro}. Others have extended these frameworks to price-maker \acp{vpp}~\cite{wang2024two} and low-carbon scheduling~\cite{zhu2025low}, simultaneously quantifying uncertainties in generation, load, and prices.

While the aforementioned studies focus primarily on energy trading, \acp{vpp} can also leverage flexible assets to provide ancillary services, enhancing system stability and their revenue~\cite{yang2023optimal, shang2025uncertainty, feng2025optimal, du2025optimal, siqin2025two, esfahani2024stochastic, nokandi2023three, zamani2016day, nemati2025segan, nemati2025single, nemati2026integration}. Yang et al.~\cite{yang2023optimal} utilized non-linear programming for joint participation in energy, reserve, and regulation markets, characterizing uncertainties via confidence intervals. Shang et al.~\cite{shang2025uncertainty} applied an interval probability method to manage volatility across emergency levels in frequency modulation markets. Feng et al.~\cite{feng2025optimal} proposed a chance-constrained \ac{dro} framework defining interval ambiguity sets for \ac{res} generation. Du et al.~\cite{du2025optimal} modeled the physical-economic feasible region of \acp{vpp} within a two-stage optimization for joint markets, and Siqin et al.~\cite{siqin2025two} addressed simultaneous energy, reserve, and carbon trading. Additionally, Esfahani et al.~\cite{esfahani2024stochastic} focused on frequency regulation from aggregated small \ac{ess}, employing an \ac{so}--\ac{ro} approach with \ac{cvar} to manage capacity and price risks. Nokandi et al.~\cite{nokandi2023three} developed a three-stage \ac{so} model incorporating intraday demand response exchanges, while Zamani et al.~\cite{zamani2016day} applied point estimate methods for joint electrical and thermal scheduling. Li et al.~\cite{li2026risk} introduce a risk-aware \ac{so} framework for multi-market bidding that broadens the scope of uncertainty to encompass not only \ac{res} variability but also the compliance willingness of users. Robust frameworks have also been refined; Nemati et al. proposed methods capturing asymmetric uncertainties in sequential markets~\cite{nemati2025segan}, handling non-linear couplings in simultaneous markets via flexible \ac{ro}~\cite{nemati2025single}, and integrating concentrated solar power through reformulated two-stage \ac{ro}~\cite{nemati2026integration}.

As discussed above, the domain of hourly multi-market participation is quite mature, with sophisticated optimization techniques handling diverse uncertainties. However, the transition of electricity markets toward higher temporal resolution leaves existing hourly models insufficient for capturing intra-hourly dynamics. Limited research has been focused on addressing this shift~\cite{li2025internal, yuanyuan2023distributionally, wang2023optimal}. Li et al.~\cite{li2025internal} proposed a \ac{dro} strategy for a 96-period market, but they simplified market prices to hourly resolution to aid convergence. Yuanyuan et al.~\cite{yuanyuan2023distributionally} focused on a 96-period declaration strategy with flexible ramping products, yet their uncertainty modeling was limited to wind power. Wang et al.~\cite{wang2023optimal} utilized Italian market data to model hourly \ac{dam} and 15-minute reserve markets but did not fully account for intra-hourly price and demand volatility. Consequently, there remains a critical gap for a holistic approach that enables \ac{vpp} participation in high-resolution energy and reserve markets while comprehensively accounting for the intra-hourly exposure to uncertainties in market prices, \ac{res} generation, and demand. {\color{black}In this context, Table~\ref{table:Literature} contrasts the key features of this study against the reviewed literature.}

{\color{black}To fill the identified gaps and align with the transition of electricity markets worldwide toward sub-hourly resolution,} this paper proposes a novel bidding strategy for \ac{rvpp} participation in the \ac{dam} and \ac{srm}. The core contribution is the development of an \ac{mbro} framework capable of simultaneously managing the multiple uncertainties inherent in electricity prices, \ac{ndrs} production, and demand. The concept of \ac{mbro} was developed in~\cite{busing2012new} to mitigate the inherent conservatism of classical \ac{ro} theory~\cite{bertsimas04}. The proposed \ac{mbro} framework in this paper fundamentally differs from the traditional \ac{ro} approach by explicitly modeling intra-hourly uncertainty exposure. Standard \ac{ro} formulations typically apply a single global robustness budget across the time horizon, often resulting in a binary outcome where specific time periods are fully protected against maximum possible deviations—leading to excessive conservatism—while remaining periods are treated optimistically with no protection once the budget is exhausted. In contrast, the proposed \ac{mbro} discretizes the uncertainty spectrum into distinct "bounds" corresponding to different deviation magnitudes. The \ac{rvpp} operator can, in this way, define flexible robustness budgets for each bound rather than relying on a single global parameter. By assigning higher budgets to lower bounds (representing frequent, moderate deviations) and lower budgets to upper bounds (representing rare, extreme deviations), the formulation ensures broader protection across more time intervals without forcing the solution to accommodate the maximum protection for every period. The aim is to effectively balance risk and profitability to yield practical bidding decisions tailored to high-resolution markets. The study further details the mathematical proof for the linear reformulation of this bi-level \ac{mbro} problem and validates the approach through extensive comparisons with classic \ac{ro} and hourly-resolution benchmarks.

\begin{table*}[t!]
  \centering
  \caption{Comparison of the approach in this paper and recent literature.}
\scriptsize
  \setlength{\tabcolsep}{2pt}
  \renewcommand{\arraystretch}{1.25}
  \vspace{-1em}
  \begin{threeparttable}
  \begin{tabular}{@{}cccccccccccc@{}}  % 10 columns
    \toprule
    & \multicolumn{2}{c}{\textbf{Market}}
    & \multicolumn{2}{c}{\textbf{Market resolution}}
    & \multicolumn{3}{c}{\textbf{Uncertainty}}
    & \multicolumn{1}{c}{\textbf{Intra-hourly}}
    & \multicolumn{2}{c}{\textbf{Uncertainty treatment}}
    & \multicolumn{1}{c}{\textbf{Method}}
   \\
   \cmidrule(lr){2-3}\cmidrule(lr){4-5}\cmidrule(lr){6-8}\cmidrule(lr){10-11}
   \textbf{Ref.} & \textbf{Energy} & \textbf{Reserve}
    & \textbf{24-period} & \textbf{96-period}
    & \textbf{Price} & \textbf{\ac{res}} & \textbf{Demand}
    & \textbf{uncertainty}
    & \textbf{Single-bound} & \textbf{Multi-bound}
    & \textbf{}
    \\
    \cmidrule{1-12}

    \cite{liu2025two}
    & $\bullet$  &  
    & $\bullet$ &  
    &  & $\bullet$ &
    &
    & $\bullet$  & 
    &  \ac{ro}
    \\ [0.2em]

    \cite{wu2024two, feng2024optimal}
    & $\bullet$ &  
    & $\bullet$ &  
    &  & $\bullet$ &
    &
    & $\bullet$  & 
    &  \ac{dro}
    \\ [0.2em]

    \cite{li2023robust}
    & $\bullet$ &  
    & $\bullet$ &  
    &  & $\bullet$ & $\bullet$
    &
    & $\bullet$  & 
    &  \ac{ro}
    \\ [0.2em]

    \cite{falabretti2023scheduling}
    & $\bullet$  &  
    & $\bullet$ &  
    &  & $\bullet$ & $\bullet$
    &
    &  & 
    &  \ac{so}
    \\ [0.2em]

    \cite{cao2025scenario}
    & $\bullet$ &  
    & $\bullet$ &  
    &  & $\bullet$ & $\bullet$
    &
    & $\bullet$  & 
    &  \ac{dro}
    \\ [0.2em]

    \cite{wang2024optimal}
    & $\bullet$ &  
    & $\bullet$ &  
    & $\bullet$ &  &
    &
    & $\bullet$  & 
    &  \ac{cvar}--\ac{dro}
    \\ [0.2em]

    \cite{mei2024two, xiao2024windfall}
    & $\bullet$  &  
    & $\bullet$ &  
    & $\bullet$  & $\bullet$ &
    &
    &  & 
    &  \ac{so}
    \\ [0.2em]

    \cite{shafiekhani2022optimal}
    & $\bullet$ &  
    & $\bullet$ &  
    & $\bullet$ & $\bullet$ &
    &
    & $\bullet$  & 
    &  \ac{igdt}
    \\ [0.2em]

    \cite{afzali2025flexibility}
    & $\bullet$ &  
    & $\bullet$ &  
    & $\bullet$ & $\bullet$ &
    &
     & $\bullet$  & 
    &  \ac{so}--\ac{ro}
    \\ [0.2em]

    \cite{ma2025data, zhu2025low}
    & $\bullet$ &  
    & $\bullet$ &  
    & $\bullet$ & $\bullet$ & $\bullet$
    &
    & $\bullet$  & 
    &  \ac{ro}
    \\ [0.2em]

    \cite{wang2024two}
    & $\bullet$ &  
    & $\bullet$ &  
    & $\bullet$ & $\bullet$ & $\bullet$
    &
    & $\bullet$  & 
    &  \ac{dro}
    \\ [0.2em]

    \cite{yang2023optimal, shang2025uncertainty}
    & $\bullet$ & $\bullet$
    & $\bullet$ &  
    &  & $\bullet$ &
    &
    &  & 
    &  Probability distribution
    \\ [0.2em]

    \cite{feng2025optimal}
    & $\bullet$ & $\bullet$
    & $\bullet$ &  
    &  & $\bullet$ &
    &
    &  & 
    &  Chance constrained \ac{dro}
    \\ [0.2em]

    \cite{du2025optimal}
    & $\bullet$ & $\bullet$
    & $\bullet$ &  
    &  & $\bullet$ & $\bullet$
    &
    &  & 
    &  Chance constrained
    \\ [0.2em]

    \cite{siqin2025two}
    & $\bullet$ & $\bullet$
    & $\bullet$ &  
    & $\bullet$ & $\bullet$ &
    &
    &  & 
    &  \ac{dro}
    \\ [0.2em]

    \cite{esfahani2024stochastic}
    & $\bullet$ & $\bullet$
    & $\bullet$ &  
    & $\bullet$ &  & $\bullet$
    &
    & $\bullet$  & 
    &  \ac{so}--\ac{ro}
    \\ [0.2em]

    \cite{nokandi2023three, zamani2016day, li2026risk}
    & $\bullet$ & $\bullet$
    & $\bullet$ &  
    & $\bullet$ & $\bullet$ & $\bullet$
    &
    &  & 
    &  \ac{so}
    \\ [0.2em]

    \cite{nemati2025segan, nemati2025single, nemati2026integration}
    & $\bullet$ & $\bullet$
    & $\bullet$ &  
    & $\bullet$ & $\bullet$ & $\bullet$
    &
    & $\bullet$  & 
    &  \ac{ro}
    \\ [0.2em]

    \cite{li2025internal}
    & $\bullet$ &  
    &  & $\bullet$
    &  & $\bullet$ &
    &
    & $\bullet$  & 
    &  \ac{dro}
    \\ [0.2em]

    \cite{yuanyuan2023distributionally}
    & $\bullet$ & $\bullet$
    &  & $\bullet$
    &  & $\bullet$ &
    &
    & $\bullet$  & 
    &  \ac{dro}
    \\ [0.2em]

    \cite{wang2023optimal}
    & $\bullet$ & $\bullet$
    &  & $\bullet$
    &  & $\bullet$ &
    &
    & $\bullet$  & 
    &  \ac{ro}
    \\ [0.2em]

    \textbf{This paper}
    & $\bullet$ & $\bullet$
    &  & $\bullet$
    & $\bullet$ & $\bullet$ & $\bullet$
    & $\bullet$
    &  & $\bullet$
    & \ac{mbro}
    \\  

    \bottomrule
  \end{tabular}
  \end{threeparttable}
  \label{table:Literature}
\end{table*}
%\end{sidewaystable}

\subsection{Paper Contributions}

The main contributions of this paper are outlined as follows:

\begin{itemize}

\item \textbf{Proposes an \ac{mbro} bidding framework for \ac{rvpp} market participation:} This approach simultaneously handles multiple uncertainties related to electricity prices, \ac{ndrs} production, and demand consumption. Unlike classical \ac{ro}, which often yields overly conservative solutions by treating all uncertainty realizations uniformly, the proposed \ac{mbro} formulation provides a more realistic representation of uncertainty. By distinguishing between frequent, moderate deviations and rare, extreme ones, the approach generates less conservative, more implementable schedules that enhance the profitability of the \ac{rvpp}.

\item \textbf{Addresses intra-hourly uncertainty exposure in market transitions:} {\color{black}The model aligns with the recent transition of electricity markets from hourly to finer temporal bidding structures.} The study highlights that ignoring intra-hour uncertainty leads to a systematic misestimation of imbalance risks by smoothing out intra-hour volatility in \ac{rvpp} trading and scheduling. This paper provides more realistic and granular bidding decisions for modern electricity markets by handling uncertainty at a quarter-hourly time resolution.

\item \textbf{Quantifies the value of multi-bound robustness and granularity:} A rigorous comparative analysis against standard \ac{ro} benchmarks is conducted across different uncertainty-handling strategies. It also quantifies the economic and technical impact of market time-resolution (15-min vs. hourly) on decision-making, offering insights into the trade-offs between computational complexity, uncertainty coverage, and profit maximization in the new market era.

\end{itemize}

The remainder of this paper is organized as follows. Section~\ref{sec:Problem_description} defines the operational scope and problem statement. The mathematical formulation of the proposed \ac{mbro} strategy for the \ac{rvpp}'s joint participation in the \ac{dam} and \ac{srm} is derived in Section~\ref{sec:Formulation}. A comprehensive numerical analysis and discussion of the results are presented in Section~\ref{sec:Case_Studies}. Finally, Section~\ref{sec:Conclusion} summarizes the key findings and suggests potential directions for future research.

\section{Problem Description}
\label{sec:Problem_description}

\begin{figure}[t!]
    \centering
        {\includegraphics[width=\linewidth]{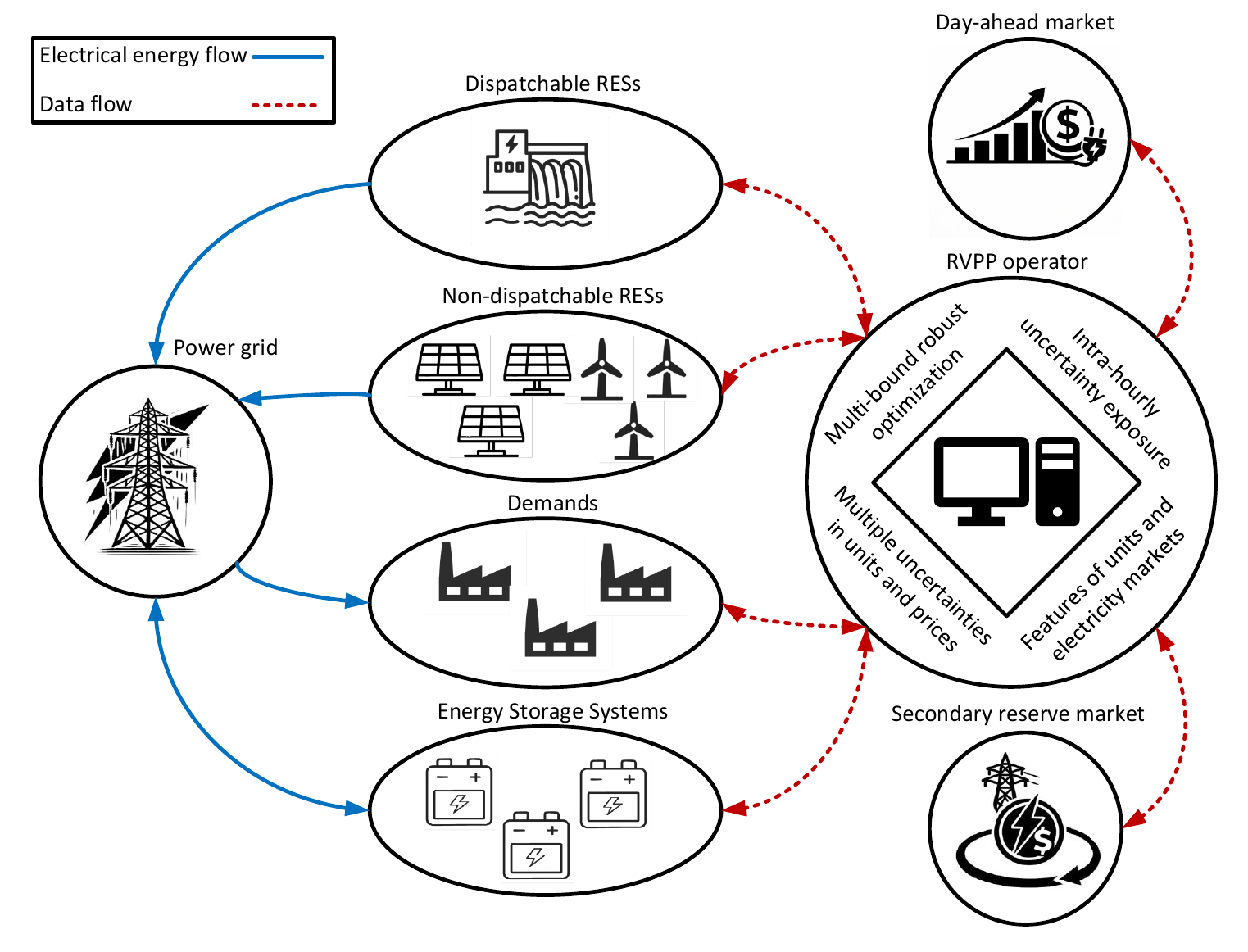}}
    \caption{The operational framework of the proposed RVPP.}
    \label{fig:RVPP_Layout}
    %\vspace{3em}
\end{figure}

The operational framework of the proposed \ac{rvpp} is conceptually illustrated in Figure~\ref{fig:RVPP_Layout}. As depicted, the \ac{rvpp} functions as a central aggregator that coordinates a diverse portfolio of assets, including \ac{drs} (e.g., hydro plant), \ac{ndrs} (e.g., wind and solar \ac{pv}), demand, and \ac{ess}. The figure distinguishes between two primary interaction flows essential for the \ac{rvpp}'s operation. First, the electrical energy flow (blue lines) represents the physical power exchange between the power grid and the internal assets. The \ac{rvpp} facilitates the grid integration of \ac{ndrs} by balancing their variability with controllable dispatchable units and storage, with the aim to ensure technical feasibility and reliable demand satisfaction. Second, the data and decision flow (red dashed lines) represents the decision-making core where the \ac{rvpp} operator gathers technical data and availability forecasts from internal units while simultaneously monitoring market conditions. Based on this information, the operator submits optimal bids to the \ac{dam} for energy trading and the \ac{srm} for ancillary services. {\color{black}In this framework, the \ac{rvpp} is modeled as a price-taker and submits zero-price bids to ensure dispatch. The rationale for this assumption lies in the \ac{rvpp}'s relatively small size compared to the grid size, as well as the economic characteristics of its aggregated assets, which are primarily \ac{ndrs} with low marginal operating costs.} Once the markets clear, the operator disaggregates the commitments, communicating specific dispatch set-points back to the internal units to maintain the schedule.

%Regarding the bidding structure used in this thesis, it is assumed, for modeling purposes, that the \gls{rvpp} bids at zero or very low prices such that the bid is always cleared. This simplification is justified by the relatively small size of the \gls{rvpp} in the context of the overall system and by the fact that most of its units are \glspl{ndrs} with low marginal operating costs. For resources within the \gls{rvpp} that possess non-negligible production or opportunity costs, this assumption implies they are treated as price-takers, focusing the analysis on quantity dispatch decisions rather than strategic price formation. In practice, market participants must respect competition law and market monitoring rules, and offering energy at prices that are systematically below cost or value could be considered anti-competitive; however, a detailed analysis of compliant bidding strategies is beyond the scope of this thesis. Accordingly, the \gls{rvpp} is considered to be in a price-taker position, meaning its bids do not influence the market-clearing price. The proposed bidding strategy for \gls{rvpp} in this thesis does not use the supply-demand balancing constraints to determine the electricity price. Indeed, it is assumed that the unknown electricity prices are input forecast parameters in the proposed model. This assumption is valid since knowing the full network configuration is not reasonable for any \gls{rvpp} operator.

A critical challenge for the \ac{rvpp} operator is managing the multiple uncertainties in units and prices highlighted in the optimization block of Figure~\ref{fig:RVPP_Layout}. To maximize profitability while ensuring robustness, the operator must account for stochasticity in \ac{ndrs} production, demand consumption, and market prices. While recent literature suggests that flexible \ac{ro} allows operators to adjust conservatism via uncertainty budgets~\cite{ma2025data, zhu2025low, wang2023optimal, nemati2025segan}, standard \ac{ro} approaches often treat all deviations uniformly. This can lead to overly conservative schedules that fail to capture the nuances of more granular bidding structures. To address these limitations, this study proposes an \ac{mbro} approach, mathematically formulated in the following section. Unlike classical methods, \ac{mbro} segments the uncertainty spectrum into distinct bounds (e.g., frequent small deviations vs. rare extreme events). This granular handling of uncertainty allows the \ac{rvpp} to prioritize frequent deviations over rare sharp ones, resulting in less conservative bidding strategies that maintain computational efficiency and are better aligned with the high-resolution requirements of modern energy and reserve markets.

\section{Formulation}
\label{sec:Formulation}
This section details the mathematical framework of the proposed \ac{mbro} approach for the \ac{rvpp} operation. \Cref{subsec:Deterministic_Formulation} first outlines the deterministic model for participation in the \ac{dam} and \ac{srm} in line with existing literature~\cite{ma2025data, zhu2025low, wang2023optimal, nemati2025segan, nemati2025single}. Subsequently, to effectively address the inherent volatility of electricity prices, generation from \ac{ndrs}, and demand consumption, \Cref{subsec:Multi-Bound_Robust} introduces the proposed \ac{mbro} framework as a novel contribution of this work.

\subsection{Deterministic Problem}
\label{subsec:Deterministic_Formulation}

\subsubsection{Objective Function}
\label{subsubsec:Objective_Function}

The objective function defined in~\eqref{RVPP: Obj_Deterministic} serves to maximize the \ac{rvpp}'s total operational profit derived from participation in both the \ac{dam} and the \ac{srm}. This formulation aggregates revenues from energy trading and bidirectional reserve provision, subtracting the operational expenditures associated with \acp{drs}, \acp{ndrs}, and \acp{ess}. %To ensure physical feasibility while maintaining computational tractability for decision-making, the formulation models essential short-term operations at a quarter-hourly dispatch resolution $\Delta t$.

\begingroup
\allowdisplaybreaks
\vspace{-.5em}
\begin{align} \label{RVPP: Obj_Deterministic}
&\mathop {\max}\limits_{{\Xi}^{F}} \sum\limits_{t \in \mathscr{T}} { {\lambda_t^{DA}p_t^{DA}\Delta t   } } + \sum\limits_{t \in \mathscr{T}} {\left[ {{\lambda _t^{{SR, \uparrow}}r_t^{SR,\uparrow} }  +{\lambda _t^{{SR, \downarrow}}r_t^{SR,\downarrow} }  } \right]}
 \!- \sum\limits_{t \in \mathscr{T}} {\!\sum\limits_{c \in \mathscr{C}} { {{C_c}p_{c,t}\Delta t %+C_{c}^{SU} v_{c,t}^{SU} +C_{c}^{SD} v_{c,t}^{SD}
   - \sum\limits_{t \in \mathscr{T}} {\!\sum\limits_{r \in \mathscr{R}} {C_rp_{r,t}\Delta t} }
  \!- \sum\limits_{t \in \mathscr{T}} {\!\sum\limits_{s \in \mathscr{S}} {C_{s}p_{s,t}\Delta t} } 
}  }}  
\end{align}
\vspace{.1em}
\endgroup

\subsubsection{Supply--Demand Constraints}
\label{subsubsec:Supply_Demand_Constraints}

The fundamental requirement for the \ac{rvpp} to meet its market obligations is the preservation of power balance at every time step. The supply--demand constraint~\eqref{cons:Supply-Demand1} is thus defined to ensure that the \ac{rvpp}'s aggregated output matches its commitment regardless of whether secondary reserves are activated in the upward or downward direction, or remain inactive. To implement this, the model employs the reserve state vectors, $\boldsymbol{r}_{t}^{SR}=\{r_{t}^{SR,\uparrow},-r_{t}^{SR,\downarrow},0\}$ for the \ac{rvpp}, and $\boldsymbol{r}_{u,t}=\{r_{u,t}^{\uparrow},-r_{u,t}^{\downarrow},0\}$ for individual assets $u\in\mathscr{U}$, including \acp{ndrs}, \acp{ess}, \acp{drs}, and demands. Thus, equation~\eqref{cons:Supply-Demand1} describes a system of three distinct balance equations that fully capture the deterministic response required for each activation scenario~\cite{nemati2025single}.

\begingroup
\allowdisplaybreaks
\vspace{-.8em}
%\begin{subequations}
\begin{align}
    &\!\sum\limits_{r \in \mathscr{R}} \left[ p_{r,t} \!+ \boldsymbol{r}_{r,t} \right] \!+ \!\sum\limits_{s \in \mathscr{S}} \left[ p_{s,t} \!+ \boldsymbol{r}_{s,t} \right] \!+ \!\sum\limits_{c \in \mathscr{C}} \left[ p_{c,t} \!+ \boldsymbol{r}_{c,t} \right] %+\sum\limits_{s \in \mathscr{S}} \left[ p_{s,t} + \boldsymbol{r}_{s,t} \right]
     - \sum\limits_{d \in \mathscr{D}} \left[ p_{d,t} - \boldsymbol{r}_{d,t} \right] = p_{t}^{DA}+ \boldsymbol{r}_{t}^{SR}~;
    & 
    \forall t \label{cons:Supply-Demand1}
\end{align}
%\label{RVPP:Supply-Demand}
%\end{subequations}
\vspace{.1em}
\endgroup

%\vspace{1em}
\subsubsection{Dispatchable Unit Constraints}
\label{subsubsec:DRES_Constraints}

In this study, \ac{drs} units are defined as renewable energy assets that can be operated in a controllable manner, similar to conventional thermal units. These resources include units such as hydro plants and can be scheduled by the \ac{rvpp} within their operational limits to provide both energy and reserve. To model this controllability, constraints~\eqref{Deterministic: DRES1} and~\eqref{Deterministic: DRES2} define the operational limits for energy and reserve provision, contingent on the binary commitment variable $v_{c,t}$. Constraint~\eqref{Deterministic: DRES3} ensures compliance with daily energy production limits, often dictated by seasonal water management policies (e.g., for hydro plants). Finally, temporal constraints regarding minimum up and down times are omitted here for brevity but follow the standard \ac{milp} formulation presented in~\cite{carrion2006computationally}.

\begingroup
\allowdisplaybreaks
\begin{subequations}
\vspace{-.9em}
\begin{align}
    &p_{c,t} + r_{c,t}^{\uparrow} \leq \bar P_{c} v_{c,t}~; 
    &
    \forall c, t \label{Deterministic: DRES1} \\
    & \ubar P_{c} v_{c,t} \leq p_{c,t} - r_{c,t}^{\downarrow}~; 
    &
    \forall c, t \label{Deterministic: DRES2} \\
    &\sum\limits_{t \in \mathscr{T}} \left[{p_{c,t}\Delta t} + r_{c,t}^{\uparrow} \right] \le \bar E_{c}~;
    &
    \forall c \label{Deterministic: DRES3} 
    \end{align}    
\label{Deterministic: DRES}
\end{subequations}
\vspace{-.8em}
\endgroup

\subsubsection{Non-dispatchable Unit Constraints}
\label{subsubsec:RES_Constraints}

The category of \ac{ndrs} includes stochastic assets like wind and solar \ac{pv}, where the maximum output is mostly determined by meteorological forecasts rather than operator dispatch. Despite this, they can participate in the \ac{srm} by modulating curtailment levels: downward reserve is provided by reducing output, whereas upward reserve is available when the unit is pre-curtailed below its available potential~\cite{zhang2022frequency, yin2021state}. In the model, Constraint~\eqref{cons: NDRES1} caps production at the forecasted limit $P_{r,t}$, while Constraint~\eqref{cons: NDRES2} defines the lower feasibility bounds for simultaneous energy and reserve schedules.
\begingroup
\allowdisplaybreaks
\begin{subequations}
%\vspace{-.8em}
\begin{align}
    & p_{r,t}+r_{r,t}^{\uparrow} \leq P_{r,t}~; 
    & 
    \forall r,t \label{cons: NDRES1} \\ 
    & \ubar P_{r} \le p_{r,t}-r_{r,t}^{\downarrow}~; 
    & 
    \forall r,t  \label{cons: NDRES2} 
    \end{align}
\label{RVPP: NDRES}
\end{subequations}
\vspace{-1em}
\endgroup

\subsubsection{Demand Constraints}
\label{subsubsec:Demand_Constraints}

Electrical demand represents a source of both flexibility and stochasticity within the \ac{rvpp} portfolio~\cite{feng2025optimal}. The feasible operating domain, which accounts for simultaneous power consumption and reserve provision, is established by constraints~\eqref{cons: Demand1} and~\eqref{cons: Demand2}, where the lower consumption limit is governed by forecast uncertainty. The minimum energy requirement over the scheduling horizon is enforced by constraint~\eqref{cons: Demand3}.
\begingroup
%\vspace{-.5em}
\allowdisplaybreaks
\begin{subequations}
\begin{align}
    & P_{d,t} \le p_{d,t} - r_{d,t}^{\uparrow}~;
    &
    \forall d,t  \label{cons: Demand1} \\
    & p_{d,t} + r_{d,t}^{\downarrow} \le \bar P_{d}~;
    & 
    \forall d,t  \label{cons: Demand2} \\
    & \ubar E_{d} \le \sum\limits_{t \in \mathscr{T}} \left[{p_{d,t}\Delta t} - r_{d,t}^{\uparrow} \right]~;
    &
    \forall d \label{cons: Demand3}
\end{align}
\label{RVPP: Demand}
\end{subequations}
\vspace{-.5em}
\endgroup

\subsubsection{ESS Constraints}
\label{subsubsec:ES_Operation}

 \acp{ess} are rapidly gaining popularity as flexibility provision assets, due to their capability to efficiently mitigate the intermittency of \acp{ndrs}, among other interesting features. The mathematical framework governing \ac{ess} operation is presented in~\eqref{Deterministic: ES}, enabling simultaneous optimization of energy arbitrage and reserve provision~\cite{nemati2026integration}. Constraints~\eqref{Deterministic: ES2}--\eqref{Deterministic: ES5} establish the operational boundaries for \ac{ess} power and reserve, ensuring that sufficient capacity is maintained for both upward and downward reserve provision in both charging $(+)$ and discharging $(-)$ states. The binary variable $v_{s,t}$ is utilized to explicitly differentiate between the charging and discharging states, thus preventing simultaneous operation. The resulting net power output and total reserve capacity are aggregated in~\eqref{Deterministic: ES6}--\eqref{Deterministic: ES8}. Regarding energy management, equation~\eqref{Deterministic: ES9} tracks the state of charge dynamics, while constraint~\eqref{Deterministic: ES10} imposes a daily cyclic balance by equating the initial and final stored energy. Note that this constraint can be relaxed according to \ac{rvpp} operator needs. The allocation of storage capacity for upward and downward reserves is governed by coefficients $\sigma_{s}^{\uparrow}$ and $\sigma_{s}^{\downarrow}$ in~\eqref{Deterministic: ES11}--\eqref{Deterministic: ES12}. Finally, constraint~\eqref{Deterministic: ES13} adjusts the feasible energy limits to account for the energy allocated to reserve capacity.
\begingroup
\allowdisplaybreaks
\begin{subequations}
\begin{align}
    &\ubar {P}_{s}^{+} v_{s,t} \leq p_{s,t}^{+} - r_{s,t}^{+,\uparrow}~; 
    &\forall s, t \label{Deterministic: ES2} \\
    & p_{s,t}^{+} + r_{s,t}^{+,\downarrow} \leq \bar {P}_{s}^{+} v_{s,t}~;
    &\forall s, t \label{Deterministic: ES3} \\
    &p_{s,t}^{-} + r_{s,t}^{-,\uparrow} \leq \bar {P}_{s}^{-} \left( 1 - v_{s,t} \right)~; 
    &\forall s, t \label{Deterministic: ES4} \\
    &\ubar {P}_{s}^{-} \left( 1 - v_{s,t} \right) \leq p_{s,t}^{-} - r_{s,t}^{-,\downarrow}~; 
    &\forall s, t \label{Deterministic: ES5} \\
    &p_{s,t} = p_{s,t}^{-} - p_{s,t}^{+}~; 
    &\forall s, t \label{Deterministic: ES6} \\
    &r_{s,t}^{\uparrow} = r_{s,t}^{+,\uparrow} + r_{s,t}^{-,\uparrow}~; 
    &\forall s, t \label{Deterministic: ES7} \\
    &r_{s,t}^{\downarrow} = r_{s,t}^{+,\downarrow} + r_{s,t}^{-,\downarrow}~; 
    &\forall s, t \label{Deterministic: ES8} \\
    & e_{s,t} = e_{s,t-1} + p_{s,t}^{+} \eta_s^{+} \Delta t - \frac{p_{s,t}^{-} \Delta t}{\eta_{s}^{-}}~;
    &\forall s, t \backslash\{1\} \label{Deterministic: ES9} \\
    & e_{s,1} = e_{s, |\mathscr{T}|}~;% = \alpha_s, 
    &\forall s \label{Deterministic: ES10} \\
    & \sum_{t \in \mathscr{T}} \frac{r_{s,t}^{\uparrow} \Delta t} {\eta_s^{-}} \leq \sigma_{s}^{\uparrow} \left( \bar E_s - \ubar E_s \right)~; 
    &\forall s \label{Deterministic: ES11} \\
    & \sum_{t \in \mathscr{T}} r_{s,t}^{\downarrow} \eta_s^{+} \Delta t \leq \sigma_{s}^{\downarrow} \left( \bar E_s - \ubar E_s \right)~; 
    &\forall s \label{Deterministic: ES12} \\
    &\ubar E_s + \sigma_s^{\downarrow} \left( \bar E_s - \ubar E_s \right) \leq e_{s,t} \leq \bar E_{s} - \sigma_{s}^{\downarrow} \left( \bar E_{s} - \ubar E_{s} \right);& 
    \forall s, t \label{Deterministic: ES13}   
    %
    %& v_{s,t} \in \{0,1\}; 
    %&\forall s, t \label{Deterministic: ES14}
\end{align}
\label{Deterministic: ES}
\end{subequations}
%\vspace{-.6em}
\endgroup
\subsubsection{Reserve Provision Constraints}
\label{subsubsec:Reserve_Constraints}

Although the energy production limits are technology-specific, the mathematical framework governing reserve provision is unified across the \ac{rvpp} portfolio. Constraints~\eqref{Deterministic: Reserve1} and~\eqref{Deterministic: Reserve2} determine the permissible upward and downward reserve capacity for each asset $u\in\mathscr{U}$ that comprises the \ac{rvpp} portfolio. This reserve capacity is strictly bounded by their ramp-rate characteristics and the secondary reserve activation time. Constraints~\eqref{Deterministic: Reserve3} and~\eqref{Deterministic: Reserve4} cap the reserve contribution based on the unit's maximum electrical rating and the reserve provision limit factors assigned by the \ac{rvpp} operator~\cite{nemati2025single}.
\begingroup
\allowdisplaybreaks
%\vspace{-1em}
\begin{subequations}
\begin{align}
    & r_{u,t}^{\uparrow} \le T^{SR} R_{u}^{\uparrow}~;
    & 
    \forall u,t  \label{Deterministic: Reserve1} \\
    & r_{u,t}^{\downarrow} \le T^{SR} R_{u}^{\downarrow}~;
    & 
    \forall u,t  \label{Deterministic: Reserve2} \\
    & r_{u,t}^{\uparrow} \le \beta^{\uparrow}_u \bar P_{u}~;
    & 
    \forall u,t  \label{Deterministic: Reserve3} \\
    & r_{u,t}^{\downarrow} \le \beta^{\downarrow}_u \bar P_{u}~;
    & 
    \forall u,t  \label{Deterministic: Reserve4}
    \end{align}    
\label{Deterministic: Reserve}
\vspace{-2em}
\end{subequations}
\endgroup
\subsection{Multi-Bound Robust Model}
\label{subsec:Multi-Bound_Robust}

The deterministic formulation in \Cref{subsec:Deterministic_Formulation} assumes perfect foresight, neglecting the inherent volatility of market prices and asset energy outputs. However, uncertainty in electricity and reserve prices, as well as variations in \ac{ndrs} generation and demand, can significantly impact the \ac{rvpp}'s financial performance. To mitigate these risks, the optimization problem is extended to an \ac{mbro} framework in~\Cref{subsubsec:Bilevel_formulation}. This is formulated as a bi-level problem: the first level maximizes the \ac{rvpp}'s objective function analogous to the deterministic model, while the second level minimizes the \ac{rvpp}'s profit by identifying the worst-case realization of uncertain parameters within defined bounds. Unlike traditional \ac{ro}, which typically defines a single interval for uncertainty (often leading to overly conservative solutions dominated by extreme scenarios), the \ac{mbro} framework segments the uncertainty spectrum into distinct bounds (e.g., deviation levels). By assigning specific budgets to these separate bounds, the model captures a more granular profile of risk. {\color{black}To solve the resulting bi-level problem efficiently with standard optimization tools, the inner worst-case (second-level) problem is reformulated into an equivalent tractable form, enabling a single-level \ac{milp} implementation.} \Cref{subsubsec:Inner_Reformulation} details this reformulation of the second-level uncertainty problem. Finally, \Cref{subsubsec:MILP_Formulation} presents the final \ac{milp} formulation of the \ac{mbro} model as a single-level problem solvable by commercial \ac{milp} solvers.

\subsubsection{Bi-Level Formulation}
\label{subsubsec:Bilevel_formulation}

The bi-level \ac{mbro} problem for the \ac{rvpp}'s participation in the \ac{dam} and \ac{srm} is formulated in~\eqref{Uncertainty: Obj}--\eqref{Uncertainty_ALL}. The objective function~\eqref{Uncertainty: Obj} seeks to maximize the base deterministic profit (represented by the first two lines, which account for expected market revenues based on price forecasts and unit operational costs) while explicitly subtracting the worst-case revenue loss caused by electricity price uncertainty in the third line. The set of first-level decision variables, $\Xi^F$, remains consistent with the deterministic model presented in~\Cref{subsec:Deterministic_Formulation}. The worst-case realization of market prices is determined by the inner maximization problem, which is performed over the sets of worst-case time periods for \ac{dam} and \ac{srm} prices ($\mathscr{T}^{DA}_k, \mathscr{T}^{SR,\uparrow}_k, \mathscr{T}^{SR,\downarrow}_k$). The cardinality of these sets is equal to the value of the assigned uncertainty budgets ($\Gamma^{DA}_k$, $\Gamma^{SR,\uparrow}_k$, $\Gamma^{SR,\downarrow}_k$). These budgets are defined over the 96 quarter-hourly intervals of the daily horizon and constrain the number of time periods in which the uncertain parameter is allowed to deviate to the level associated with a specific bound $k$. Crucially, to ensure logical consistency, the summation of budgets across all bounds for a given parameter is limited to 96. This constraint enforces mutual exclusivity, ensuring that any specific time period can be assigned to at most one deviation level. This flexibility allows the operator to shape a less conservative uncertainty distribution—for example, by assigning larger budgets to frequent, minor deviations and restricting extreme deviations to a few rare instances. The uncertainty in \ac{dam} electricity prices is characterized by asymmetric intervals, expressed as ${\lambda}_t^{DA} {\in} \left[\tilde{\lambda}_t^{DA}-\check{\lambda}^{DA}_{k,t}, \tilde{\lambda}_t^{DA} + \hat{\lambda}^{DA}_{k,t} \right]$, where the magnitudes of positive and negative deviations, $\hat{\lambda}^{DA}_{k,t}$ and $\check{\lambda}^{DA}_{k,t}$, are generally distinct. Within these bounds, the determination of the worst-case price is contingent on the \ac{rvpp}'s net position in the market: a negative deviation constitutes the worst-case scenario when the portfolio is selling energy, whereas a positive deviation represents the worst case during energy purchase intervals. Conversely, for secondary reserve prices, only negative deviations are relevant to the robust analysis, as formulated by the intervals $\left({\lambda}_t^{{SR, \uparrow}} {\in} \left[ {\bar{\lambda}}_t^{{SR, \uparrow}}-\check{\lambda}_{k,t}^{{SR, \uparrow}}, {\bar{\lambda}}_t^{{SR, \uparrow}} \right]\right.$ and $\left.{\lambda}_t^{{SR, \downarrow}} {\in} \left[ {\bar{\lambda}}_t^{{SR, \downarrow}}-\check{\lambda}_{k,t}^{{SR, \downarrow}},{\bar{\lambda}}_t^{{SR, \downarrow}} \right]\right)$. By focusing on these adverse deviations only, the model ensures that the optimization problem accurately reflects the worst-case conditions defined by the chosen uncertainty budgets, ignoring favorable deviations that would increase the \ac{rvpp}'s profit.
\begingroup
\allowdisplaybreaks
\begin{align} 
&\mathop {\max}\limits_{{\Xi^{F}}} \scalebox{1.5}{\Bigg\{} \sum\limits_{t \in \mathscr{T}} {\left[ {\tilde{\lambda}_t^{DA}p_t^{DA}\Delta t +{{\bar{\lambda}}_t^{{SR, \uparrow}}r_t^{SR,\uparrow} } +{{\bar{\lambda}}_t^{{SR, \downarrow}}r_t^{SR,\downarrow} }  } \right]}  \nonumber \\& - \sum\limits_{t \in \mathscr{T}} {\!\sum\limits_{r \in \mathscr{R}} {C_rp_{r,t}\Delta t} } \!- \sum\limits_{t \in \mathscr{T}} {\!\sum\limits_{s \in \mathscr{S}} {C_{s}p_{s,t}\Delta t} } \!- \sum\limits_{t \in \mathscr{T}} \!\sum\limits_{c \in \mathscr{C}} { {{C_c}p_{c,t}\Delta t } }  \nonumber
\\ & - \mathop{\max}\limits_{\substack{
    \left\{
    \substack{
    \mathscr{T}^{DA}_k, \mathscr{T}^{SR,\uparrow}_k, \mathscr{T}^{SR,\downarrow}_k 
    }
    \right\}
    }}
    \left\{ \sum\limits_{t \in {\mathscr{T}^{DA}_k}} \sum_{k \in \mathscr{K}} {\check{\lambda}_{k,t}^{DA} {y^{DA}_{k,t}} } + \sum\limits_{t \in {\mathscr{T}^{SR,\uparrow}_k}} \sum_{k \in \mathscr{K}} {\check {\lambda}_{k,t}^{{SR, \uparrow}}r_t^{SR,\uparrow} } + \sum\limits_{t \in {\mathscr{T}^{SR,\downarrow}_k}} \sum_{k \in \mathscr{K}} {\check {\lambda}_{k,t}^{{SR, \downarrow}}r_t^{SR,\downarrow} }  
\right\}   \scalebox{1.5}{\Bigg\}} 
\label{Uncertainty: Obj}
\end{align}
\endgroup

The constraints governing the proposed bi-level \ac{mbro} framework are detailed in~\eqref{Uncertainty_ALL}. Constraint~\eqref{Uncertainty: Asymetric_DAprice} addresses the asymmetric characteristics of \ac{dam} price uncertainty through the positive auxiliary variable ${y^{DA}_{k,t}}$. By distinguishing between positive (${\hat{\lambda}_{k,t}^{DA}}$) and negative (${\check{\lambda}_{k,t}^{DA}}$) price deviations, this variable effectively models the worst-case price realization contingent on the direction of the \ac{rvpp}'s energy trade in each period~\cite{nemati2026integration}. The uncertainty associated with \ac{ndrs} generation and demand consumption is managed by equations~\eqref{Uncertainty: NDRES} and~\eqref{Uncertainty: Demand}, respectively. These constraints incorporate specific uncertainty bounds to account for scenarios that worsen the objective function: \ac{ndrs} production is modeled with a focus on potential negative deviations $\left({P}_{r,t} \in [{\bar{P}}_{r,t}-\check{P}_{r,k,t}, {\bar{P}}_{r,t} ]\right)$, whereas demand is modeled to account for potential positive deviations $\left({P}_{d,t} \in [{\ubar{P}}_{d,t},{\ubar{P}}_{d,t}+\hat{P}_{d,k,t}]\right)$. The worst-case realizations for these parameters are determined by inner-level maximization problems performed over the sets $\mathscr{T}_{r,k}$ and $\mathscr{T}_{d,k}$, whose cardinalities, as discussed earlier, are equal to the uncertainty budgets $\Gamma_{r,k}$ and $\Gamma_{d,k}$ for each bound $k$. Consequently, constraints~\eqref{Uncertainty: NDRES} and \eqref{Uncertainty: Demand} apply the respective deviations (i.e., to reduce $\check{P}_{r,k,t^\prime}$ and/or to increase $\hat{P}_{d,k,t^\prime}$) to the forecast reference only if the index $t^\prime$ matches the current period $t$ and simultaneously belongs to the selected worst-case set (i.e., ${t^\prime = t}$ and ${t^\prime \in \mathscr{T}_{r,k}}$ or ${t^\prime \in \mathscr{T}_{d,k}}$).  Finally, all deterministic constraints from~\Cref{subsec:Deterministic_Formulation} that remain unchanged under uncertainty are integrated in~\eqref{Uncertainty: Other_Cons_Det}.
\begingroup
\allowdisplaybreaks
\begin{subequations}
\begin{align}
    &- \frac{\check{\lambda}_{k,t}^{DA}}{\hat{\lambda}_{k,t}^{DA}} {y^{DA}_{k,t}} \le {p_t}^{DA} \Delta t \le {y^{DA}_{k,t}}~; 
    \hspace{17.7em}   
    &\forall k, t \in \mathscr{K}, \mathscr{T} \label{Uncertainty: Asymetric_DAprice} \\
    & p_{r,t}+r_{r,t}^{\uparrow} \leq {\bar{P}}_{r,t} - \mathop {\max}\limits_{ \left\{ \mathscr{T}_{r,k} \right\} } \left\{ \sum_{t^\prime \in \mathscr{T}_{r,k}, t^\prime = t } \, \sum_{k \in \mathscr{K}} \check{P}_{r,k,t^\prime} \right\}~;  
    & \forall r,t \in \mathscr{R},\mathscr{T} \label{Uncertainty: NDRES} \\ 
    & p_{d,t} - r_{d,t}^{\uparrow} \geq {\ubar{P}}_{d,t} + \mathop {\max}\limits_{ \left\{ \mathscr{T}_{d,k}  \right\} } \left\{ \sum_{t^\prime \in \mathscr{T}_{d,k}, t^\prime = t} \, \sum_{k \in \mathscr{K}} \hat{P}_{d,k,t^\prime}  \right\}~; 
    & \forall d,t  \in \mathscr{D},\mathscr{T} \label{Uncertainty: Demand}\\
%
    %& {y^{DA}_{k,t}} \geq 0~; 
    %& \forall k, t \in \mathscr{K}, \mathscr{T}  \label{Uncertainty: Auxillary_price}\\
    %
    &\eqref{cons:Supply-Demand1}, \eqref{Deterministic: DRES}, \eqref{cons: NDRES2}, \eqref{cons: Demand2}\text{--}\eqref{cons: Demand3}, \eqref{Deterministic: ES}, \eqref{Deterministic: Reserve}~;  
    &   
    \label{Uncertainty: Other_Cons_Det}
\end{align}
\label{Uncertainty_ALL}
\end{subequations}
\endgroup
\subsubsection{Inner Problems Reformulation}
\label{subsubsec:Inner_Reformulation}

The maximization term in the third line of the objective function~\eqref{Uncertainty: Obj} (the protection function) conceptually identifies the worst-case scenario via time sets $\mathscr{T}^{DA}_k, \mathscr{T}^{SR,\uparrow}_k, \mathscr{T}^{SR,\downarrow}_k$; however, since these sets are not predetermined, their selection logic must be recast as an equivalent linear optimization problem to facilitate the derivation of the final robust formulation~\cite{bertsimas04}. {\color{black}Specifically, set membership is represented through positive auxiliary variables in the linear problem, whose optimal values yield the same worst-case choice as the original set-based definition by limiting their allowable deviations across time periods and within each bound.} To achieve this, by treating the optimal values of the upper-level variables (${y^{DA^*}_{k,t}}$, $r_t^{SR,\uparrow^{*}}$, $r_t^{SR,\downarrow^{*}}$) as fixed parameters in the lower-level problem and applying Proposition 1 (see Appendix), the protection function in~\eqref{Uncertainty: Obj} is transformed into the linear problem~\eqref{Protection_Function_obj}. In this formulation, constraints~\eqref{Protection_obj: con1}--\eqref{Protection_obj: con3} restrict the summation of each of the positive auxiliary variables ${z^{DA}_{k,t}}$, ${z^{SR,\uparrow}_{k,t}}$, and ${z^{SR,\downarrow}_{k,t}}$ over all time periods for each uncertainty bound $k$ to the corresponding uncertainty budgets $\Gamma^{DA}_k$, $\Gamma^{SR,\uparrow}_k$, and $\Gamma^{SR,\downarrow}_k$, respectively. Simultaneously, constraints~\eqref{Protection_obj: con4}--\eqref{Protection_obj: con6} ensure that the summation of these auxiliary variables across all bounds for any given time period does not exceed 1. This linear structure mathematically replicates the mutual exclusivity of the worst-case sets, ensuring that the optimal value of~\eqref{Protection_obj: obj} matches the protection function in~\eqref{Uncertainty: Obj}. The dual variables associated with each constraint are defined alongside these equations and are subsequently utilized in~\Cref{subsubsec:MILP_Formulation} to derive the final \ac{milp} formulation.
\begingroup
\allowdisplaybreaks
\begin{subequations}
\begin{align}
    & %\beta^{O} = %\nonumber \\ & 
    {\max} \left\{         
        \sum\limits_{t \in {\mathscr{T}}} \sum_{k \in \mathscr{K}} {\check{\lambda}_{k,t}^{DA} {y^{DA^*}_{k,t}} {z^{DA}_{k,t}} } + \sum\limits_{t \in {\mathscr{T}}} \sum_{k \in \mathscr{K}} {\check {\lambda}_{k,t}^{{SR, \uparrow}} r_t^{SR,\uparrow^{*}} {z^{SR,\uparrow}_{k,t}} } + \sum\limits_{t \in {\mathscr{T}}} \sum_{k \in \mathscr{K}} {\check {\lambda}_t^{{SR, \downarrow}} r_t^{SR,\downarrow^*} {z^{SR,\downarrow}_{k,t}} } \right\}
    \label{Protection_obj: obj}
    \end{align}
    \begin{align}
    \nonumber\text{st.} \\
    & \sum_{t \in {\mathscr{T}}} {z^{DA}_{k,t}} \le \Gamma_k^{DA}: \phi_k^{DA}~;   
    & \forall {k \in {\mathscr{K}}} \label{Protection_obj: con1} \\
    & \sum_{t \in {\mathscr{T}}} {z^{SR,\uparrow}_{k,t}} \le \Gamma_k^{SR,\uparrow}: \phi_k^{SR,\uparrow}~; 
    & \forall {k \in {\mathscr{K}}} \label{Protection_obj: con2} \\
    & \sum_{t \in {\mathscr{T}}} {z^{SR,\downarrow}_{k,t}} \le \Gamma_k^{SR,\downarrow}: \phi_k^{SR,\downarrow}~;    
    & \forall {k \in {\mathscr{K}}} \label{Protection_obj: con3} \\
    & 0 \le \sum_{k \in {\mathscr{K}}} {z^{DA}_{k,t}} \le 1: \zeta_{t}^{DA}~; 
    & \forall {t \in {\mathscr{T}}}   \label{Protection_obj: con4} \\
    & 0 \le \sum_{k \in {\mathscr{K}}} {z^{SR,\uparrow}_{k,t}} \le 1: \zeta_{t}^{SR,\uparrow}~; 
    & \forall {t \in {\mathscr{T}}} \label{Protection_obj: con5} \\
    & 0 \le \sum_{k \in {\mathscr{K}}} {z^{SR,\downarrow}_{k,t}} \le 1: \zeta_{t}^{SR,\downarrow}~; 
    & \forall {t \in {\mathscr{T}}} \label{Protection_obj: con6}
\end{align}
\label{Protection_Function_obj}
\end{subequations}
\endgroup

The protection function describing the uncertainty of \ac{ndrs} electrical production in constraint~\eqref{Uncertainty: NDRES} (the maximization term on the right-hand side) is recast as the linear formulation~\eqref{Protection_Function_NDRES} by applying Proposition 2 (see Appendix). To achieve the same worst-case selection logic as in~\eqref{Uncertainty: NDRES}, the optimization problem~\eqref{Protection_Function_NDRES} is defined individually for each time period $t$ and \ac{ndrs} unit $r$~\cite{nemati2026integration}. The objective function~\eqref{Protection_NDRES: obj} maximizes the production deviation for the current period $t^\prime = t$. Constraint~\eqref{Protection_NDRES: con1} ensures that the sum of the auxiliary variable $z_{r,k,(t^\prime=t)}$ for the current period and the aggregate optimal values of variables from other protection functions ($z_{r,k,t^\prime}^{*}$ where $t^\prime \neq t$) remains within the limit $\Gamma_{r,k}$. Note that the variables $z_{r,k,t^\prime}^{*}$ correspond to other protection functions (where $t^\prime \neq t$) and are therefore treated as parameters in this constraint. Furthermore, the summation of the auxiliary variable $z_{r,k,(t^\prime=t)}$ across all uncertainty bounds is limited to 1 by constraint~\eqref{Protection_NDRES: con2}. The equivalent linear formulation for the demand uncertainty protection function in constraint~\eqref{Uncertainty: Demand} follows an identical derivation and is omitted here for brevity.

\begingroup
\allowdisplaybreaks
\begin{subequations}
\begin{align}
    & %\beta_{r, t}^{C} =
    {\max} \sum_{k \in \mathscr{K}} \check{P}_{r,k,(t^\prime=t)} \; z_{r,k,(t^\prime=t)}~; & \forall r, t \in \mathscr{R}, \mathscr{T}      
    &   
    \label{Protection_NDRES: obj} \\
    \nonumber\text{st.} \\
    & z_{r,k,(t^\prime=t)} + \sum_{t^\prime \neq t} z_{r,k,t^\prime}^{*} \le \Gamma_{r,k}: \phi_{r,k}~; 
    &   
    \forall {k \in {\mathscr{K}}} \label{Protection_NDRES: con1} \\
    & 0 \le \sum_{k \in \mathscr{K}} z_{r,k,(t^\prime=t)} \le 1: \zeta_{r,t}~; 
    & 
    \label{Protection_NDRES: con2}
\end{align}
\label{Protection_Function_NDRES}
\end{subequations}
\endgroup

\subsubsection{MILP Formulation}
\label{subsubsec:MILP_Formulation}

By leveraging strong duality theory~\cite{floudas1995nonlinear}, the inner linear maximization problems defined in~\eqref{Protection_Function_obj} and~\eqref{Protection_Function_NDRES} (as well as the analogous formulation for demand uncertainty) are replaced by their respective dual problems. Substituting the dual forms directly into the objective function~\eqref{Uncertainty: Obj} and constraints~\eqref{Uncertainty: NDRES} and~\eqref{Uncertainty: Demand} yields the final single-level \ac{milp} formulation presented in~\eqref{Single-level_MILP}. In this formulation, the objective function~\eqref{MILP: Obj} retains the deterministic profit terms in its first two lines, while the third line explicitly penalizes the objective function for the worst-case revenue losses arising from \ac{dam} and \ac{srm} price uncertainties. Constraint~\eqref{MILP_asymetric1} replicates the asymmetric price logic from~\eqref{Uncertainty: Asymetric_DAprice}, and constraints~\eqref{MILP: DAMprice}--\eqref{MILP: downSRMprice} enforce the dual restrictions corresponding to the price uncertainty problem defined in~\eqref{Protection_Function_obj}.

The worst-case limits for \ac{ndrs} production are governed by constraints~\eqref{MILP: NDRES1}--\eqref{MILP: NDRES3}. Specifically, constraint~\eqref{MILP: NDRES1} defines the robust production upper bound. To model the discrete logical condition where the remaining budget term ($\Gamma_{r,k} - \sum_{t^\prime \neq t} z_{r,k,t^\prime}^{*}$) only corresponds to 0 or 1, the binary variable $\chi_{r,k,t}$ is introduced. Consequently, the positive auxiliary variable $y_{r,k,t}$ is established in~\eqref{MILP: NDRES1} and bounded via constraint~\eqref{MILP: NDRES2} using the Big-M method~\cite{floudas1995nonlinear}, thus linearizing the dual term $\sum_{k \in \mathscr{K}} \chi_{r,k,t} \phi_{r,k} + {\zeta}_{r,t}$. To ensure logical consistency, constraint~\eqref{MILP: NDRES4} enforces mutual exclusivity, ensuring only one uncertainty bound is active per time period, while constraint~\eqref{MILP: NDRES5} ensures the total number of active worst-case periods for each bound $k$ matches the assigned uncertainty budget. Constraint~\eqref{MILP: NDRES3} represents the standard dual constraints derived from the \ac{ndrs} protection function~\eqref{Protection_Function_NDRES}. Similarly, constraints~\eqref{MILP: Demand1}--\eqref{MILP: Demand3} manage demand uncertainty. These follow the same structural logic as the \ac{ndrs} constraints, with the primary distinction being the direction of the worst-case deviation. The domains of all positive auxiliary variables are defined in~\eqref{MILP: Auxilliary_obj}--\eqref{MILP: Auxilliary_cons}, and the remaining deterministic constraints are consolidated in~\eqref{MILP: Other_Cons_Det}. Critically, this refined \ac{milp} framework allows for the flexible management of multiple uncertainty sources and enables the selection of distinct conservatism levels across different bounds via the uncertainty budgets.

\begingroup
\begin{subequations}
\allowdisplaybreaks
\begin{align} \label{MILP: Obj}
&\mathop {\max}\limits_{{\Xi^{F}, \Xi^{S}}} \scalebox{1.1}{\Bigg\{} \sum\limits_{t \in \mathscr{T}} {\left[ {\tilde{\lambda}_t^{DA}p_t^{DA}\Delta t +{{\bar{\lambda}}_t^{{SR, \uparrow}}r_t^{SR,\uparrow} } +{{\bar{\lambda}}_t^{{SR, \downarrow}}r_t^{SR,\downarrow} }  } \right]}  \nonumber \\& - \sum\limits_{t \in \mathscr{T}} {\!\sum\limits_{r \in \mathscr{R}} {C_rp_{r,t}\Delta t} } \!- \sum\limits_{t \in \mathscr{T}} {\!\sum\limits_{s \in \mathscr{S}} {C_{s}p_{s,t}\Delta t} } \!- \sum\limits_{t \in \mathscr{T}} \!\sum\limits_{c \in \mathscr{C}} { {{C_c}p_{c,t}\Delta t } }  \nonumber
\\
& - \sum\limits_{k \in \mathscr{K}} \left[ \Gamma^{DA}_k \phi^{DA}_k + \Gamma^{SR,\uparrow}_k \phi^{SR,\uparrow}_k + \Gamma^{SR,\downarrow}_k \phi^{SR,\downarrow}_k\right] - \sum\limits_{t \in \mathscr{T}} \left[{\zeta}^{DA}_t + {\zeta}^{SR,\uparrow}_t+ {\zeta}^{SR, \downarrow}_t\right]    \scalebox{1.1}{\Bigg\}}
\end{align}
\begin{align}
    \nonumber\text{st.} \\
    &- \frac{\check{\lambda}^{DA}_{k,t}}{\hat{\lambda}^{DA}_{k,t}} {y^{DA}_{k,t}} \le {p_t}^{DA} \Delta t \le {y^{DA}_{k,t}}~; 
    & \forall k, t \in \mathscr{K}, \mathscr{T} \label{MILP_asymetric1} \\
    &{\phi^{DA}_k} + \zeta_t^{DA} \ge  \check{\lambda}^{DA}_{k,t}y^{DA}_{k,t}~; 
    & \forall k, t \in \mathscr{K}, \mathscr{T} \label{MILP: DAMprice} \\
    &{\phi^{SR,\uparrow}_k} + \zeta_t^{SR,\uparrow} \ge ~\check{\lambda}^{SR,\uparrow}_{k,t} r_t^{SR,\uparrow}~; 
    & \forall k, t \in \mathscr{K}, \mathscr{T} \label{MILP: upSRMprice} \\
    &{\phi^{SR,\downarrow}_k} + \zeta_t^{SR,\downarrow} \ge ~\check{\lambda}^{SR,\downarrow}_{k,t} r_t^{SR,\downarrow}~; 
    & \forall k, t \in \mathscr{K}, \mathscr{T} \label{MILP: downSRMprice} \\
    & p_{r,t}+r_{r,t}^{\uparrow} \leq {\bar{P}}_{r,t} - \sum\limits_{k \in \mathscr{K}} \chi_{r,k,t} \, \phi_{r,k} - {\zeta}_{r,t} = {\bar{P}}_{r,t} - \sum\limits_{k \in \mathscr{K}} y_{r,k,t}~;  
    & \forall r,t \in \mathscr{R}, \mathscr{T} \label{MILP: NDRES1} \\  
    &\phi_{r,k} +  {\zeta}_{r,t} - M (1 - \chi_{r,k,t}) \leq y_{r,k,t} \leq M \chi_{r,k,t}~;  
    & \forall r,k,t \in \mathscr{R},\mathscr{K},\mathscr{T} \label{MILP: NDRES2} \\
    & \sum_{k\in \mathscr{K}} \chi_{r,k,t} \le 1~; 
    & \forall r, t \in \mathscr{R}, \mathscr{T} \label{MILP: NDRES4} \\
    & \sum_{t\in \mathscr{T}} \chi_{r,k,t} = \Gamma_{r,k}~; 
    & \forall r, k \in \mathscr{R}, \mathscr{K} \label{MILP: NDRES5} \\
    &\phi_{r,k} + \zeta_{r,t} \ge  \check{P}_{r,k,t}~; 
    & \forall r, k, t \in \mathscr{R},\mathscr{K},\mathscr{T} \label{MILP: NDRES3} \\
    & p_{d,t} - r_{d,t}^{\uparrow} \geq {\ubar{P}}_{d,t} + \sum\limits_{k \in \mathscr{K}} \chi_{d,k,t} \, \phi_{d,k} + {\zeta}_{d,t} = {\ubar{P}}_{d,t} + \sum\limits_{k \in \mathscr{K}} y_{d,k,t}~; 
    & \forall d,k,t \in \mathscr{D},\mathscr{K},\mathscr{T} \label{MILP: Demand1}\\
    &\phi_{d,k} +  {\zeta}_{d,t} - M (1 - \chi_{d,k,t}) \leq y_{d,k,t} \leq M \chi_{d,k,t}~; 
    & \forall d,k,t \in \mathscr{D},\mathscr{K},\mathscr{T}   \label{MILP: Demand2}\\
    & \sum_{k\in \mathscr{K}} \chi_{d,k,t} \le 1~; 
    & \forall d, t \in \mathscr{D}, \mathscr{T} \label{MILP: Demand4} \\
    & \sum_{t\in \mathscr{T}} \chi_{d,k,t} = \Gamma_{d,k}~; 
    & \forall d, k \in \mathscr{D}, \mathscr{K} \label{MILP: Demand5} \\
    &\phi_{d,k} + \zeta_{d,t} \ge  \hat{P}_{d,k,t}~; 
    & \forall d,k,t \in \mathscr{D},\mathscr{K},\mathscr{T}  \label{MILP: Demand3} \\
    &\phi^{DA}_k, \phi^{SR,\uparrow}_k, \phi^{SR,\downarrow}_k, \zeta_t^{DA}, \zeta_t^{SR,\uparrow}, \zeta_t^{SR,\downarrow}, y^{DA}_{k,t} \ge 0~; 
    & \forall k,t \in \mathscr{K},\mathscr{T} \label{MILP: Auxilliary_obj} \\
    & \phi_{r,k}, \phi_{d,k}, \zeta_{r,t}, \zeta_{d,t}, y_{r,k,t}, y_{d,k,t} \ge 0~; 
    & \forall r,d, k, t \in \mathscr{R},\mathscr{D},\mathscr{K},\mathscr{T} \label{MILP: Auxilliary_cons} \\
    %
    %& \chi_{r,k,t}, \chi_{d,k,t} \in \{0, 1\}~; 
    %\hspace{16.2em}   
    %\forall r,d,k, t \in \mathscr{R},\mathscr{D},\mathscr{K},\mathscr{T} \label{MILP: Binary_cons} 
    %
    &\eqref{cons:Supply-Demand1}, \eqref{Deterministic: DRES}, \eqref{cons: NDRES2}, \eqref{cons: Demand2}\text{--}\eqref{cons: Demand3}, \eqref{Deterministic: ES}, \eqref{Deterministic: Reserve}~;  
    &   
    \label{MILP: Other_Cons_Det}
\end{align}
\label{Single-level_MILP}
\end{subequations}
\endgroup

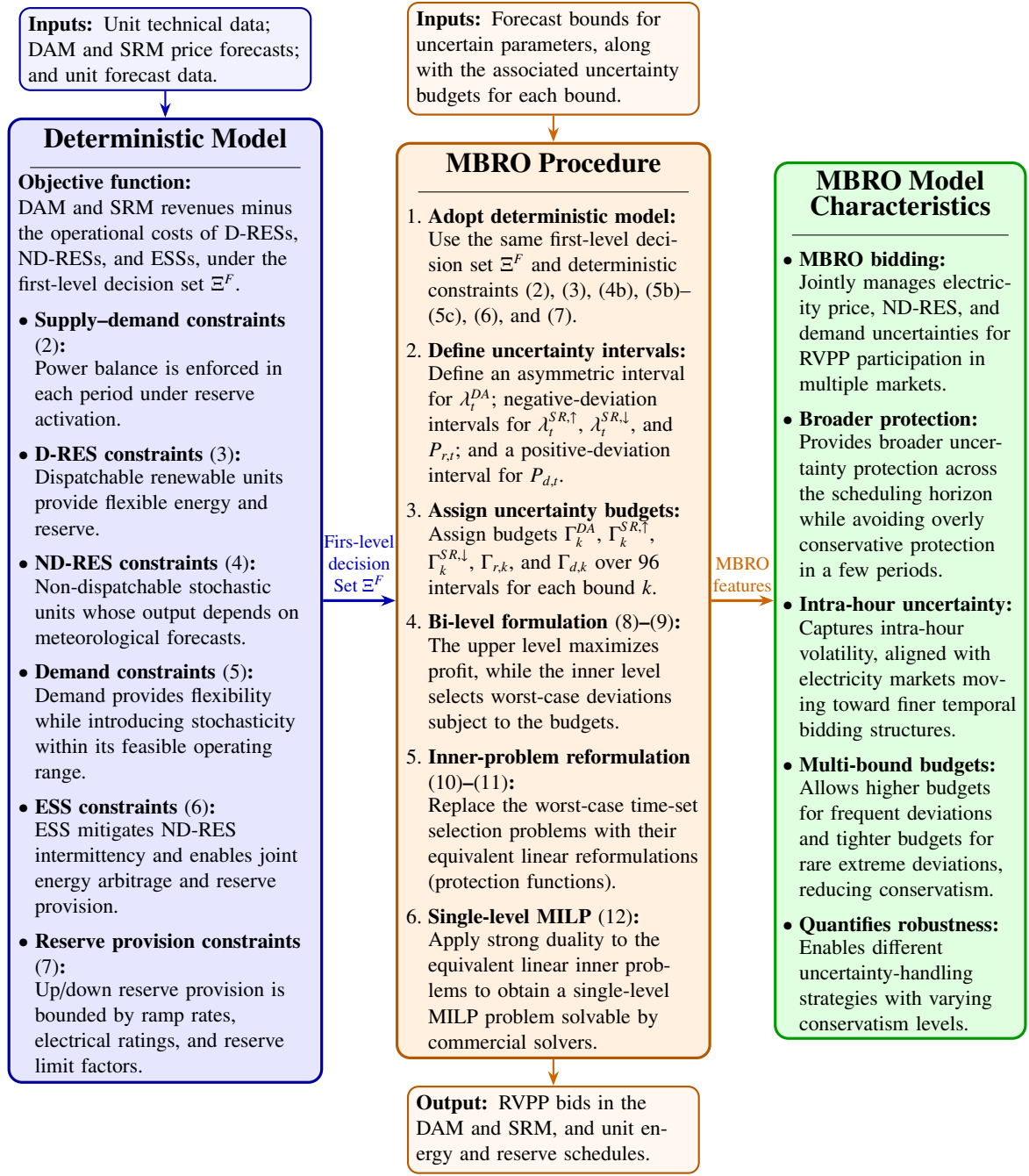
\begin{figure*}[t!]
\centering
\begin{tikzpicture}[
    font=\small,
    node distance=6mm and 8mm,
    >={Stealth[length=2.2mm]},
    % >>> Box styles <<<
    box/.style={rectangle, rounded corners=2mm, draw, very thick, align=left,
                inner sep=4.0pt, text width=0.295\textwidth},
    smallbox/.style={rectangle, rounded corners=1.6mm, draw, thick, align=left,
                     inner sep=3.6pt, text width=0.275\textwidth},
    % Part 3: 20% narrower than 'box' (0.295*0.8 = 0.236)
    box3/.style={rectangle, rounded corners=2mm, draw, very thick, align=left,
                 inner sep=3.2pt, text width=0.236\textwidth},
    title/.style={font=\bfseries, align=center},
    arr/.style={->, very thick},
    arrThin/.style={->, thick},
]

% ===================== PART 1 =====================
\node[box, fill=blue!10, draw=blue!60!black] (P1) {
  \begin{minipage}{\linewidth}
    \centering {\large\bfseries Deterministic Model}\\[-2pt]
    \rule{0.92\linewidth}{0.4pt}\\[-2pt]
    %\vspace{.1mm}

    \raggedright
    {\bfseries Objective function:}\par
    \ac{dam} and \ac{srm} revenues minus the operational costs of \acp{drs}, \acp{ndrs}, and \acp{ess}, under the first-level decision set $\Xi^F$.\par\vspace{0.8mm}

    \begin{itemize}[leftmargin=*, labelsep=0.3em, itemsep=2pt, topsep=0pt]
      \item \textbf{Supply--demand constraints \eqref{cons:Supply-Demand1}:}\\[-1pt]
            Power balance is enforced in each period under reserve activation.
      \item \textbf{\ac{drs} constraints \eqref{Deterministic: DRES}:}\\[-1pt]
            Dispatchable renewable units provide flexible energy and reserve.
      \item \textbf{\ac{ndrs} constraints \eqref{RVPP: NDRES}:}\\[-1pt]
            Non-dispatchable stochastic units whose output depends on meteorological forecasts.
      \item \textbf{Demand constraints \eqref{RVPP: Demand}:}\\[-1pt]
            Demand provides flexibility while introducing stochasticity within its feasible operating range.
      \item \textbf{ESS constraints \eqref{Deterministic: ES}:}\\[-1pt]
            \ac{ess} mitigates \ac{ndrs} intermittency and enables joint energy arbitrage and reserve provision.
      \item \textbf{Reserve provision constraints \eqref{Deterministic: Reserve}:}\\[-1pt]
            Up/down reserve provision is bounded by ramp rates, electrical ratings, and reserve limit factors.
    \end{itemize}
  \end{minipage}
};

% A small “inputs” node above Part 1
\node[smallbox, fill=blue!5, draw=blue!55!black, above=4mm of P1.north, anchor=south] (P1in) {
\textbf{Inputs:} Unit technical data; \ac{dam} and \ac{srm} price forecasts; and unit forecast data.
};

\draw[arrThin, blue!70!black] (P1in.south) -- (P1.north);

% ===================== PART 2 =====================
\node[box, fill=orange!12, draw=orange!70!black, right=11mm of P1] (P2) {
  \begin{minipage}{\linewidth}
    \centering {\large\bfseries \ac{mbro} Procedure}\\[-2pt]
    \rule{0.92\linewidth}{0.4pt}\\[-2pt]
    %\vspace{1mm}

    \begin{enumerate}[leftmargin=*, labelsep=0.3em, itemsep=2pt, topsep=0pt]
      \item \textbf{Adopt deterministic model:}\\[-1pt]
      Use the same first-level decision set $\Xi^F$ and deterministic constraints
      \eqref{cons:Supply-Demand1}, \eqref{Deterministic: DRES}, \eqref{cons: NDRES2},
      \eqref{cons: Demand2}--\eqref{cons: Demand3}, \eqref{Deterministic: ES}, and \eqref{Deterministic: Reserve}.

      \item \textbf{Define uncertainty intervals:}\\[-1pt]
      Define an asymmetric interval for ${\lambda}_t^{DA}$; negative-deviation intervals for
      ${\lambda}_t^{{SR, \uparrow}}$, ${\lambda}_t^{{SR, \downarrow}}$, and ${P}_{r,t}$; and
      a positive-deviation interval for ${P}_{d,t}$.

      \item \textbf{Assign uncertainty budgets:}\\[-1pt]
      Assign budgets $\Gamma^{DA}_k$, $\Gamma^{SR,\uparrow}_k$, $\Gamma^{SR,\downarrow}_k$,
      $\Gamma_{r,k}$, and $\Gamma_{d,k}$ over 96 intervals for each bound $k$.

      \item \textbf{Bi-level formulation \eqref{Uncertainty: Obj}--\eqref{Uncertainty_ALL}:}\\[-1pt]
      The upper level maximizes profit, while the inner level selects worst-case deviations subject to the budgets.

      \item \textbf{Inner-problem reformulation \eqref{Protection_Function_obj}--\eqref{Protection_Function_NDRES}:}\\[-1pt]
      Replace the worst-case time-set selection problems with their equivalent linear reformulations (protection functions).

      \item \textbf{Single-level \ac{milp} \eqref{Single-level_MILP}:}\\[-1pt]
      Apply strong duality to the equivalent linear inner problems to obtain a single-level \ac{milp} problem solvable by commercial solvers.
    \end{enumerate}
  \end{minipage}
};

% A small “inputs” node above Part 2
\node[smallbox, fill=orange!6, draw=orange!70!black, above=4mm of P2.north, anchor=south] (P2in) {
\textbf{Inputs:} Forecast bounds for uncertain parameters, along with the associated uncertainty budgets for each bound.
};

\draw[arrThin, orange!80!black] (P2in.south) -- (P2.north);

% A small “outputs” node below Part 2
\node[smallbox, fill=orange!6, draw=orange!70!black, below=4mm of P2.south, anchor=north] (P2out) {
\textbf{Output:} \ac{rvpp} bids in the \ac{dam} and \ac{srm}, and unit energy and reserve schedules.
};

\draw[arrThin, orange!80!black] (P2.south) -- (P2out.north);

% ===================== PART 3 (20% smaller width) =====================
\node[box3, fill=green!12, draw=green!60!black, right=9.5mm of P2] (P3) {
  \begin{minipage}{\linewidth}
    \centering {\large\bfseries \ac{mbro} Model Characteristics}\\[-2pt]
    \rule{0.92\linewidth}{0.4pt}\\[-2pt]
    %\vspace{.01mm}
    
    \begin{itemize}[leftmargin=*, labelsep=0.3em, itemsep=2pt, topsep=0pt]
      \item \textbf{\ac{mbro} bidding:}\\[-1pt]
      Jointly manages electricity price, \ac{ndrs}, and demand uncertainties for \ac{rvpp} participation in multiple markets.

      \item \textbf{Broader protection:}\\[-1pt]
      Provides broader uncertainty protection across the scheduling horizon while avoiding overly conservative protection in a few periods.

      \item \textbf{Intra-hour uncertainty:}\\[-1pt]
      Captures intra-hour volatility, aligned with electricity markets moving toward finer temporal bidding structures.

      \item \textbf{Multi-bound budgets:}\\[-1pt]
      Allows higher budgets for frequent deviations and tighter budgets for rare extreme deviations, reducing conservatism.

      \item \textbf{Quantifies robustness:}\\[-1pt]
      Enables different uncertainty-handling strategies with varying conservatism levels.
    \end{itemize}
  \end{minipage}
};

% ===================== Cross-part arrows =====================
\draw[arr, blue!70!black] (P1.east) -- node[above, align=center, font=\footnotesize]
{Firs-level\\decision\\Set $\Xi^F$} (P2.west);

\draw[arr, orange!80!black] (P2.east) -- node[above, align=center, font=\footnotesize]
{\ac{mbro}\\features} (P3.west);

\end{tikzpicture}
\caption{Procedure for developing the proposed \ac{mbro} framework and its key characteristics.}
\label{fig:MBRO_Procedure}
\end{figure*}

{\color{black}Figure~\ref{fig:MBRO_Procedure} summarizes the overall workflow of the approach presented in this section: it outlines the deterministic \ac{rvpp} formulation and the main steps leading to the \ac{mbro} model and its final single-level \ac{milp}. The figure also highlights the key characteristics enabled by the \ac{mbro} approach.} The advantages of this proposed \ac{mbro} approach—specifically its ability to yield less conservative and more realistic solutions compared to classical robust optimization—are rigorously validated through the case studies presented in~\Cref{sec:Case_Studies}.

\clearpage

\section{Case Studies}
\label{sec:Case_Studies}

This section presents and discusses simulation results obtained with the proposed \ac{mbro} framework, with the aim of assessing how multiple sources of uncertainty affect the performance of an \ac{rvpp} participating in the \ac{dam} and \ac{srm}. All experiments were executed on a Dell XPS laptop (Intel i7-1165G7, 2.8~GHz, 16~GB RAM) using CPLEX within GAMS~39.1.1. 

\subsection{System Description and \ac{mbro} Algorithm Parametrization}
\begin{figure}[t!]
    \centering
        {\includegraphics[width=\linewidth]{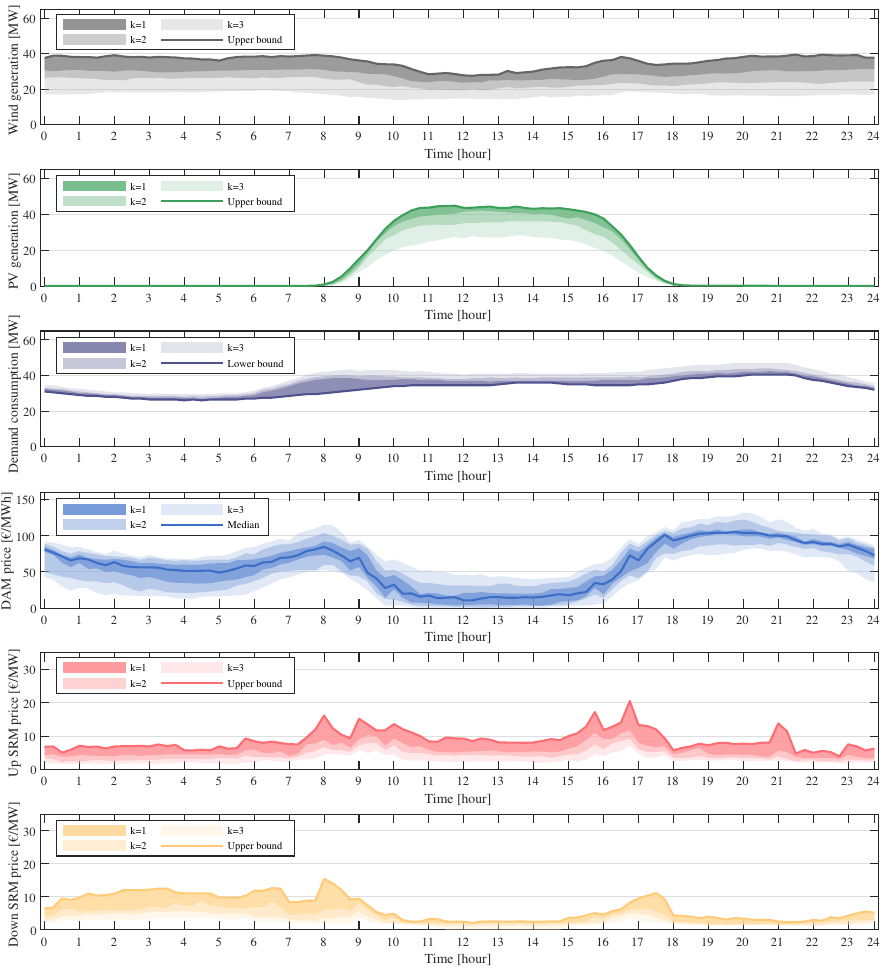}}
    \caption{Uncertainty bound in multi-bound for different uncertainty handling strategies.}
    \label{fig:Uncertainty_Bound_data}
    \vspace{2em}
\end{figure}

To account for the heterogeneity of the assets that may comprise and \ac{rvpp}, and without loss of generality, the case study considers an \ac{rvpp} located in Spain composed of a hydro plant, a wind farm, a solar \ac{pv} plant, a Li-ion \ac{ess}, and a demand. Forecast data for the uncertain inputs (namely, wind and solar \ac{pv} generation, demand consumption, \ac{dam} electricity price, and up/down \ac{srm} prices) are derived from historical records for November~2025 reported by \cite{REE2025}. {\color{black}This one-month period is used because quarter-hourly market data have only recently become available for Spain; the proposed framework is not tied to this window and can be readily applied to longer datasets as they become available.} The corresponding uncertainty bounds are illustrated in \Cref{fig:Uncertainty_Bound_data}. For each uncertain parameter, three bounds ($k=1,2,3$) are defined: $k=1$ represents the most likely deviation range around the forecast, while $k=2$ and $k=3$ capture progressively less likely (more extreme) deviations, depicted with lighter shading. {\color{black}The deviation levels and their associated budgets are user-defined robustness settings. In practice, an \ac{rvpp} operator can set them based on (i) operational experience and risk \textit{appetite} from prior market participation, and/or (ii) calibration against historical forecast-error distributions for each uncertain input (e.g., selecting deviation bounds and allocating larger budgets to more likely bounds and smaller budgets to rarer extremes). A systematic procedure to identify “best-fit” bounds/budgets for a specific portfolio is an important topic but is outside the scope of this paper.}

%The direction of the worst-case deviations is selected based on their typical impact on \ac{rvpp} profit. For wind and solar \ac{pv} production, as well as up/down \ac{srm} prices, the upper bound of the historical data is used as the deterministic reference value and downward deviations are treated as adverse realizations, since reductions in these parameters generally decrease the \ac{rvpp} revenue. For demand, the lower bound is taken as the deterministic reference and upward deviations are considered adverse, reflecting higher demand levels. For \ac{dam} prices, the deterministic reference is the median value; depending on whether the \ac{rvpp} acts as a net seller or buyer in a given period, adverse outcomes may arise from either negative or positive deviations.

The technical characteristics of the \ac{rvpp} assets are summarized in~\Cref{table:Data_Det} and are taken from \cite{nemati2025single, nemati2026integration}. \Cref{table:Data_Budget} reports the uncertainty budgets associated with the bounds in \Cref{fig:Uncertainty_Bound_data} for: (i) the proposed \ac{mbro} model with a 15-minute bidding resolution, (ii) a classic \ac{ro} model with 15-minute resolution, and (iii) an \ac{ro} model with hourly resolution. In the \ac{mbro} formulation, the sum of the budgets over all bounds ($k=1,2,3$) is constrained to lie within 0 to 96. Moreover, inner bounds are assigned larger budgets than outer bounds to reflect their higher likelihood (i.e., small deviations are more frequent than large deviations). In the \ac{ro} model with 15-minute resolution, the budget varies from 0 to 96, where larger values represent more conservative operation by allowing more periods to reach their worst-case realizations. In the hourly \ac{ro} model, the corresponding budget ranges from 0 to 24, spanning the deterministic case (0) to the most conservative case (24), in which all hours may simultaneously attain worst-case values.

Based on these budgets, four strategies are considered for the \ac{rvpp} operator: a deterministic strategy and three uncertainty-handling strategies (optimistic, balanced, and pessimistic). The following case studies are used to quantify the impact of the proposed \ac{mbro} approach, to benchmark it against the classic \ac{ro} formulation, and to examine the effect of the market time resolution (15-minute versus hourly):

\begin{itemize}
    \item \textbf{Case 1:} Evaluate the optimal operation of the \ac{rvpp} units under the proposed \ac{mbro} model for the deterministic, optimistic, balanced, and pessimistic strategies.

    \item \textbf{Case 2:} Compare \ac{rvpp} trading outcomes in the \ac{dam} and \ac{srm} using the \ac{ro} model under hourly and 15-minute market resolutions across the different bidding strategies considered.

    \item \textbf{Case 3:} Compare trading decisions and the value of different uncertainties under the classic \ac{ro} model and the proposed \ac{mbro} model for the different strategies.
\end{itemize}

\begin{table}[t!]
  \centering
   % \small
  \caption{RVPP units data.}
    \label{table:Data_Det}
  %\small
  \setlength{\tabcolsep}{1.5pt}
  \renewcommand{\arraystretch}{1.2}
 %\small
  \begin{threeparttable}
  \vspace{-.5em}
  \begin{tabular}{lccccccccccc}
    \toprule

    \multicolumn{1}{c}{\textbf{Parameter}}  
    && \multicolumn{1}{c}{\textbf{Solar PV}}
    && \multicolumn{1}{c}{\textbf{Wind farm}} 
    && \multicolumn{1}{c}{\textbf{Hydro}}
    && \multicolumn{1}{c}{\textbf{ESS}}
    && \multicolumn{1}{c}{\textbf{Demand}}
    \\

    \cmidrule{1-1} \cmidrule{3-3} \cmidrule{5-5} \cmidrule{7-7} \cmidrule{9-9} \cmidrule{11-11} 

   \multirow{1}{*}{\text{$\bar P_{u}$/$\ubar P_{u}$ [MW]}}    && 50/0 
   && 50/0 && 50/10 && - && 50/0 \\ [0.2em]

     \multirow{1}{*}{\text{$R_{u}^{\uparrow}$/$R_{u}^{\downarrow}$ [MW/min]}}    && 20/25 && 15/20 && 10/10 && 10/10 && 3/3  \\ [0.2em]

          \multirow{1}{*}{\text{$\beta_{u}^{\uparrow}$/$\beta_{u}^{\downarrow}$ [\%]}}    && 5 && 5 && 50 && 100 && 0  \\ [0.2em]

        %\multirow{1}{*}{\text{$UT_u$/$DT_u$ [hour]}}    && - && - && 1/0 && - &&  - \\ [0.2em]

        \multirow{1}{*}{\text{$\bar E_{u}/\ubar E_{u}$ [MWh]}}    && - && - && 480/- && 30/3 && -/750 \\ [0.2em]

        \multirow{1}{*}{\text{$\bar P_{s}^{+}/\bar P_{s}^{-}$ [MW]}}   && - && - && - && 10/10 && - \\ [0.2em]

        \multirow{1}{*}{\text{$\eta^{+}_s/\eta^{-}_s$ [\%]}}    && - && - && - && 95/95 && - \\ [0.2em]

        \multirow{1}{*}{\text{$C_{u}$ [€/MWh]}}    && 10 && 15 && 12.5 && 30 && -  \\ [0.2em]

\bottomrule
  \end{tabular}
\end{threeparttable}
%\vspace{1mm}
\end{table}

\begin{table}[h!]
  \centering
  \caption{Uncertainty budgets for the \ac{rvpp} operator’s strategies in different robust frameworks.}
  %\small
  \setlength{\tabcolsep}{4pt}
  \renewcommand{\arraystretch}{1}
  \vspace{-.5em}
  \begin{threeparttable}
  \begin{tabular}{lcccccccc}
    \toprule
    
    \multicolumn{1}{c}{\textbf{}}   
    && \multicolumn{3}{c}{\textbf{\ac{mbro}}} 
    && \multicolumn{1}{c}{\textbf{\ac{ro} 15-min}}
     && \multicolumn{1}{c}{\textbf{\ac{ro} hourly}}
    \\

    \multicolumn{1}{c}{\textbf{Strategy}}   
    && \multicolumn{1}{c}{\textbf{$k=1$}} 
    & \multicolumn{1}{c}{\textbf{$k=2$}}
    & \multicolumn{1}{c}{\textbf{$k=3$}} 
    && \multicolumn{1}{c}{\textbf{}}
    && \multicolumn{1}{c}{\textbf{}}
    \\

 \cmidrule{1-1} \cmidrule{3-5} \cmidrule{7-7}   \cmidrule{9-9}

    \multirow{1}{*}{Optimistic}  && 16 & 4  &  2 && 16  && 4 \\ [0.2em] 

    \multirow{1}{*}{Balanced}    && 32 & 8   & 4 && 32  && 8 \\ [0.2em]

    \multirow{1}{*}{Pessimistic}    && 48 & 12  & 6 && 48 && 12 \\ [0.2em]

\bottomrule
  \end{tabular}
\end{threeparttable}
  \label{table:Data_Budget}
    \vspace{-1em}
\end{table}

\subsection{Case 1}
\label{subsec: Case 1}

\begin{figure}[t!]
    \centering
        {\includegraphics[width=\linewidth]{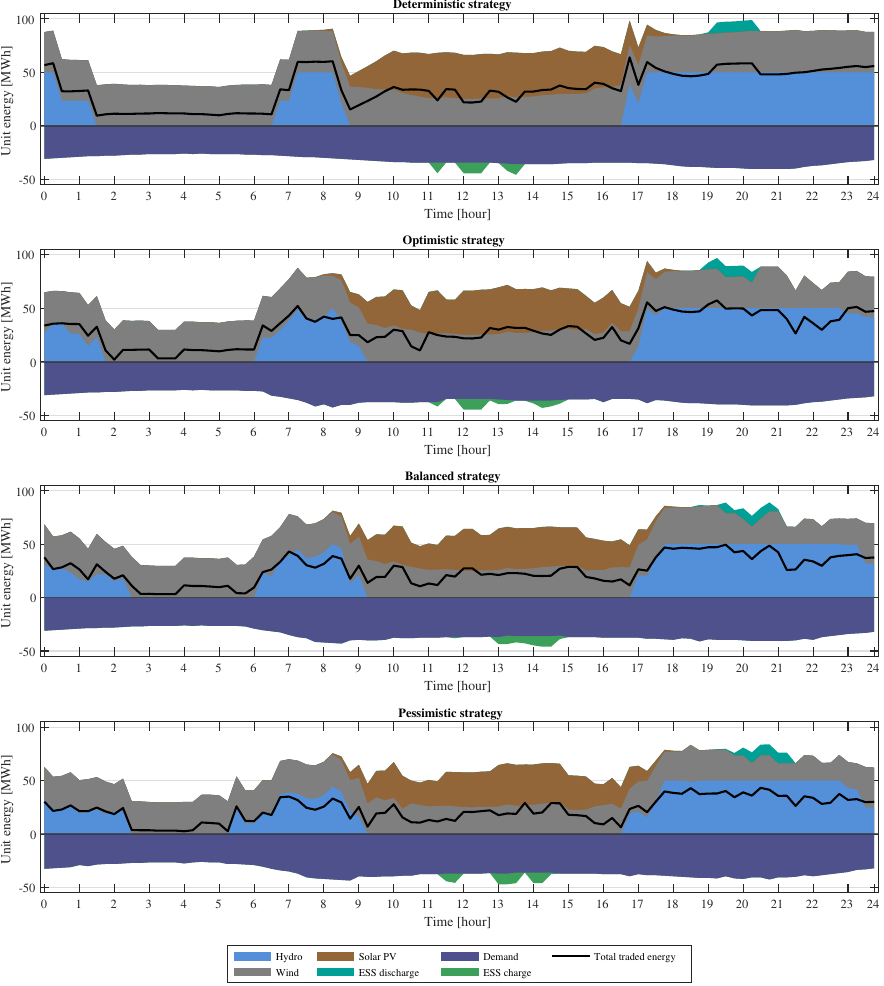}}
    \caption{Case 1 -- All RVPP unit energy in multi-bound for different uncertainty handling strategies.}
    \label{fig:ALL_stack_Case1}
    \vspace{2em}
\end{figure}

\Cref{fig:ALL_stack_Case1} shows the energy schedules of the different \ac{rvpp} units under a 15-minute market resolution for the different bidding strategies considered. In the deterministic strategy, the forecast values are treated as exact and uncertainties are neglected in the optimization problem, yielding the optimal unit dispatch under perfect-forecast assumptions. Wind and solar \ac{pv} supply energy in the \ac{dam} according to their availability. In addition, the dispatchable hydro unit mainly generates during periods with high \ac{dam} prices, i.e., around 00:00--01:30, 06:45--08:45, and 16:45--24:00 (see the electricity price in \Cref{fig:Uncertainty_Bound_data}). The \ac{ess} charges mainly between 11:15 and 13:45, when solar \ac{pv} availability is high, and discharges between 19:00 and 20:30, when \ac{dam} prices are higher. Under the optimistic strategy, uncertain parameters are allowed to deviate from their deterministic reference values according to uncertainty budgets defined in \Cref{table:Data_Budget}. Consequently, the scheduled energy of the units is affected, with lower wind and solar \ac{pv} generation and higher demand consumption in the worst case. To compensate for the reduced output of the \ac{ndrs}, the hydro plant is scheduled for more periods than in the deterministic strategy. For instance, hydro generation extends over 00:00--01:45, 06:15--09:15, and 17:00--24:00. For the more conservative balanced and pessimistic strategies, the \ac{rvpp} operator accounts for higher volatility in \ac{ndrs} when determining the unit schedules. As a result, as expected, the \ac{rvpp} trades lower energy in the \ac{dam}, with greater variability across periods. Moreover, the role of the hydro plant in mitigating \ac{ndrs} fluctuations becomes more pronounced, and this dispatchable unit is scheduled more frequently. For example, under the pessimistic strategy, hydro generation is scheduled over 00:00--02:30, 05:30--09:15, and 16:45--24:00. {\color{black}Additionally, under the pessimistic strategy, the \ac{ess} schedules charging and discharging over a wider time window than in the deterministic strategy to hedge against uncertainty. For example, the \ac{ess} charges during 11:30--14:45 and discharges during 18:45--21:30.} Overall, the results in \Cref{fig:ALL_stack_Case1} indicate that the proposed framework effectively captures the impact of uncertainty on \ac{rvpp} scheduling at a 15-minute market resolution across different uncertainty-handling strategies.

\begin{figure}[t!]
    \centering
        {\includegraphics[width=\linewidth]{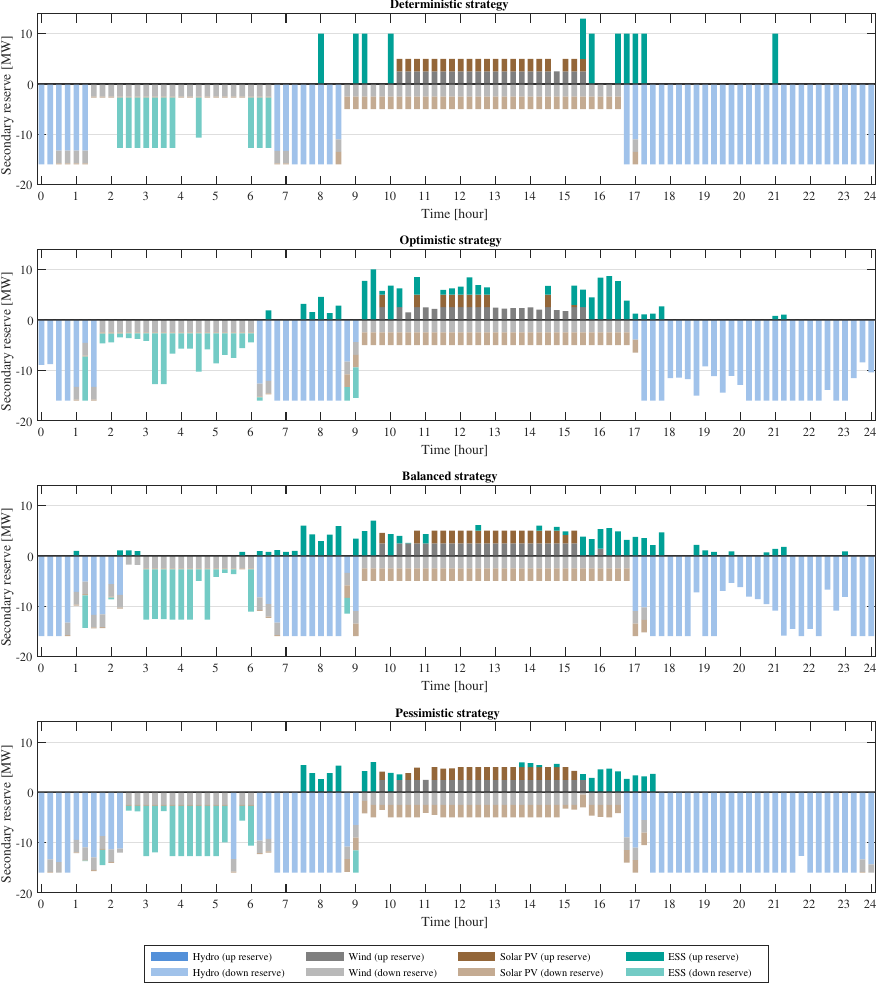}}
    \caption{Case 1 -- All RVPP unit reserve in multi-bound for different uncertainty handling strategies.}
    \label{fig:ALL_stack_reserve_Case1}
    \vspace{2em}
\end{figure}

\Cref{fig:ALL_stack_reserve_Case1} shows the shares of up and down reserve provided by the \ac{rvpp} units under a 15-minute market resolution for different bidding strategies. The results indicate that the \ac{ess} is one of the main providers of up reserve in the \ac{srm}. For instance, under the deterministic strategy, the \ac{ess} provides up reserve during several periods around 08:00--10:15, 15:30--17:30, and 21:00--21:15, when the up-\ac{srm} price is high (see \Cref{fig:Uncertainty_Bound_data}). A smaller share of up reserve is provided by the wind farm and the solar \ac{pv} plant, mainly between 10:15 and 15:45. Regarding down reserve in the deterministic strategy, most of the down reserve is supplied by the hydro plant during the hours in which it is generating and can therefore curtail its output to provide down reserve. The \ac{ess} also provides down reserve in some periods, for example between 02:15 and 06:45. In addition, the wind farm and solar \ac{pv} contribute a small share of down reserve depending on their availability to reduce production. It is worth noting that providing down reserve is generally easier than providing up reserve; accordingly, the \ac{rvpp} offers down reserve in almost all scheduling periods. In contrast, up reserve is offered in fewer periods because the \ac{rvpp} often prefers to allocate its resources to \ac{dam} energy trading rather than to up-reserve provision. Moreover, since \ac{dam} electricity prices are typically higher than \ac{srm} prices, energy trading tends to be a more reliable revenue source for the \ac{rvpp}. Nevertheless, the \ac{rvpp} still arbitrages between the two markets to enhance profitability. When adopting the optimistic strategy, and especially the more conservative balanced and pessimistic strategies, the \ac{rvpp} tends to provide up reserve over a larger number of periods, typically with a lower reserve magnitude per period compared with the deterministic strategy. 
%For example, under the balanced strategy the \ac{ess} provides up reserve mainly between 05:45 and 21:30, i.e., over more periods but with lower magnitude than in the deterministic case. 
This behaviour can be explained by the fact that, once uncertainty is explicitly considered in the proposed \ac{mbro} framework, \ac{srm} prices may be adversely affected across many periods. Consequently, the \ac{rvpp} adopts a more diversified reserve-bidding profile over time, reducing its exposure to price fluctuations in any single period and leading to a more robust and risk-averse strategy for profit maximization.

%\clearpage
\subsection{Case 2}
\label{subsec: Case 2}

\begin{figure}[t!]
    \centering
        {\includegraphics[width=\linewidth]{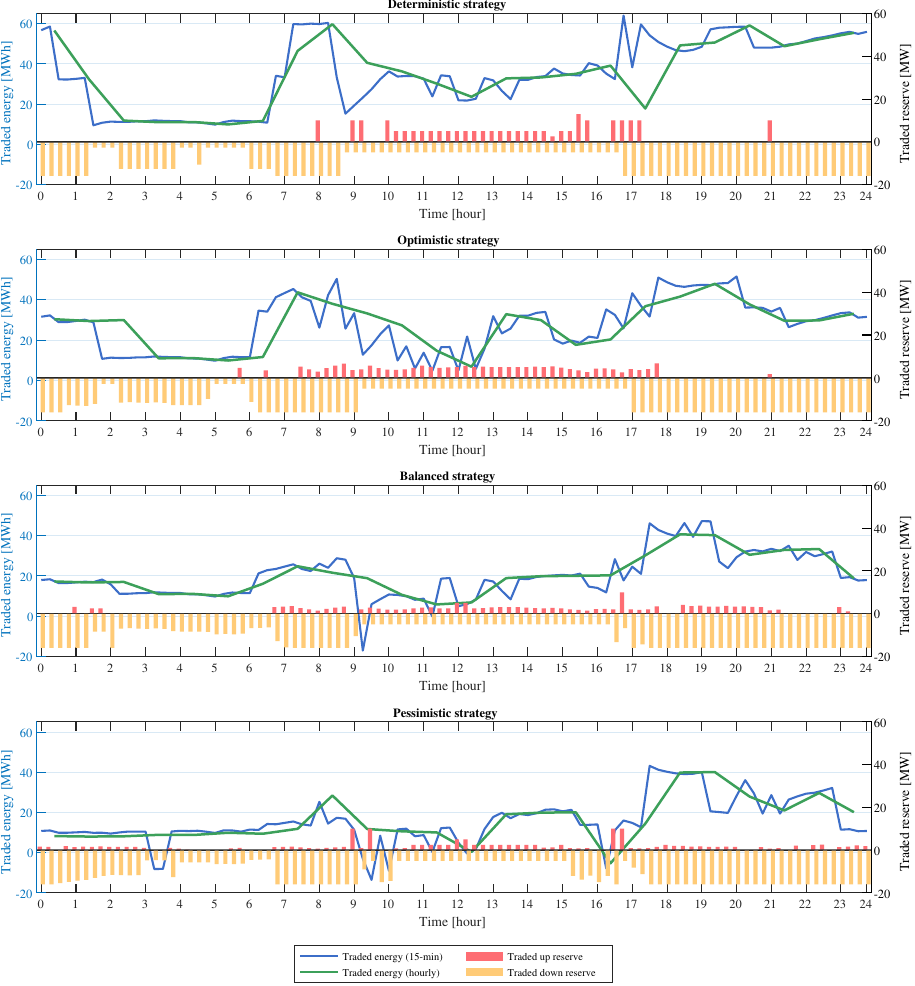}}
    \caption{Case 2 -- Total traded energy and reserve of RVPP in hourly and 15 min for different uncertainty handling strategies.}
    \label{fig:Traded_energy_Case2}
    \vspace{2em}
\end{figure}

\Cref{fig:Traded_energy_Case2} compares the \ac{rvpp} trading outcomes obtained with the \textit{classic} \ac{ro} model under hourly and 15-minute market resolutions across different uncertainty-handling strategies. Overall, the 15-minute formulation captures intra-hour trading dynamics that are inherently smoothed in the hourly framework, leading to more realistic and granular bidding decisions. 
%In particular, the \ac{rvpp} energy availability can change within the hour, so the traded energy is not necessarily the same in each of the four 15-minute intervals. For example, in the deterministic strategy, within hour 19 the traded energy between 19:00--19:15 is lower than the hourly traded value, whereas the traded energy during 19:15--20:00 is higher. 
Adopting alternative uncertainty-handling strategies can even change the trading direction in specific periods. For instance, under the balanced strategy the \ac{rvpp} becomes a net buyer during 09:15--09:30 in the 15-minute market, while the hourly model does not reflect this change. 
%Under the pessimistic strategy, such discrepancies occur more frequently; for example, during 03:15--03:45, 09:15--09:30, and 10:00--10:15 the 15-minute model schedules the \ac{rvpp} as a net buyer, whereas the corresponding hourly schedule still indicates net selling.

\Cref{table:Abs_Energy_Reserve} compares the normalized absolute differences between the hourly and 15-minute schedules of \ac{rvpp} traded energy in the \ac{dam} and traded up/down reserve in the \ac{srm}, for different uncertainty-handling strategies. For each hour, the hourly schedule is mapped to its four corresponding 15-minute periods; absolute deviations are computed period by period, summed over the full horizon, and then normalized by the total traded quantity in the hourly schedule. Larger percentages indicate stronger intra-hour variability captured by the 15-minute formulation but smoothed in the hourly formulation. In the deterministic strategy, this leads to normalized differences of 18.0\%, 64.4\%, and 15.6\% for traded energy, up reserve, and down reserve, respectively. When uncertainty is considered (optimistic, balanced, and pessimistic strategies), the normalized differences do not increase monotonically; however, a generally increasing tendency is observed toward more conservative settings. For instance, under the pessimistic strategy the normalized differences reach their maximum values of 34.2\%, 65.6\%, and 16.3\%. A key reason is that in more conservative strategies the \ac{rvpp} typically trades less energy and reserve, so a given absolute mismatch between 15-minute and hourly schedules translates into a larger normalized deviation.

\begin{table}[h!]
  \centering
  \caption{Normalized absolute differences (\%) between hourly and 15-minute traded energy in the \ac{dam} and up/down reserve in the \ac{srm}, for different uncertainty-handling strategies.}
  %\small
  \setlength{\tabcolsep}{4pt}
  \renewcommand{\arraystretch}{1}
  \vspace{-.5em}
  \begin{threeparttable}
  \begin{tabular}{lcccccc}
    \toprule
    
    \multicolumn{1}{c}{\textbf{Strategy}}   
    && \multicolumn{1}{c}{\textbf{Energy [\%]}} 
    && \multicolumn{1}{c}{\textbf{Up reserve [\%]}}
     && \multicolumn{1}{c}{\textbf{Down reserve [\%]}}
    \\

 \cmidrule{1-1} \cmidrule{3-3} \cmidrule{5-5}   \cmidrule{7-7}

    \multirow{1}{*}{Deterministic}  && 18.0  && 64.4  && 15.6 \\ [0.2em]

    \multirow{1}{*}{Optimistic}  && 20.7  && 28.7  && 10.1 \\ [0.2em] 

    \multirow{1}{*}{Balanced}    && 18.7 && 59.5  && 11.1 \\ [0.2em]

    \multirow{1}{*}{Pessimistic}    && 34.2 && 65.6 && 16.3 \\ [0.2em]

\bottomrule
  \end{tabular}
\end{threeparttable}
  \label{table:Abs_Energy_Reserve}
     \vspace{1.5em}
\end{table}

%\clearpage
\subsection{Case 3}
\label{subsec: Case 3}

\begin{figure}[t!]
    \centering
        {\includegraphics[width=\linewidth]{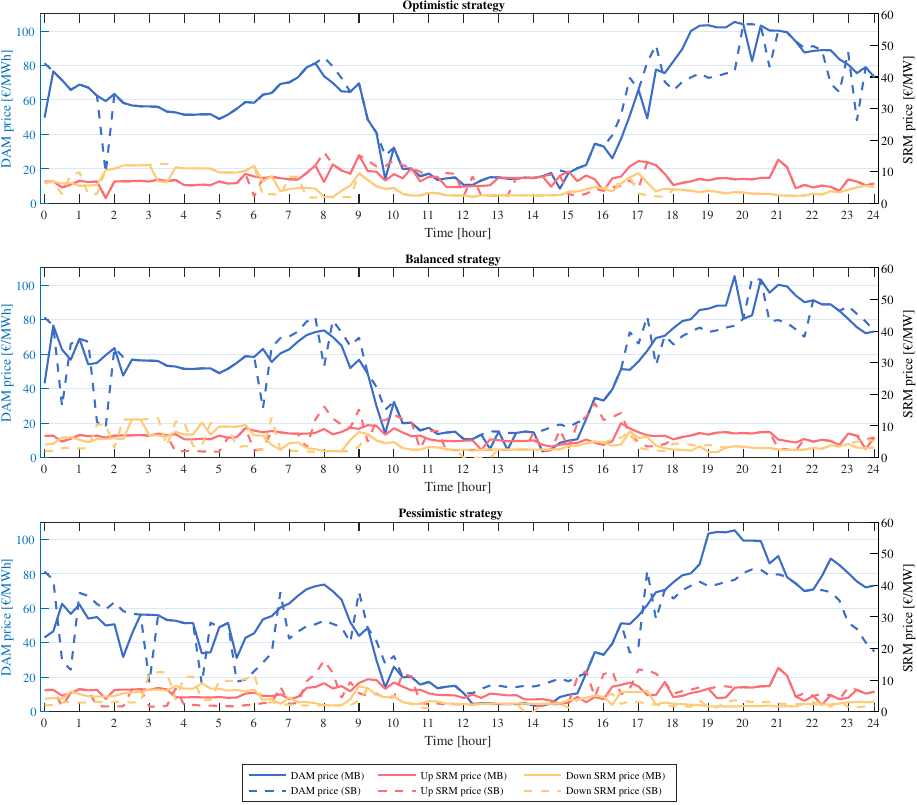}}
    \caption{Case 3 - Electricity price in multi-bound (MB) and single-bound (SB) for different uncertainty handling strategies.}
    \label{fig:Price_Case3}
    \vspace{13em}
\end{figure}

\Cref{fig:Price_Case3} compares the \ac{dam} price and the up/down \ac{srm} prices obtained with the proposed \ac{mbro} framework and the \textit{classic}, single-bound \ac{ro} model (referred to as SB in the results depicted below) under different uncertainty-handling strategies, considering 15-min market resolution. Overall, the \ac{ro} formulation produces more frequent and abrupt deviations in both \ac{dam} and \ac{srm} prices. For instance, under the optimistic strategy, the \ac{dam} price in period 1:45--2:00 deviates downward by about 70\% relative to the median value. Similar sharp negative \ac{dam} deviations appear in other periods, such as 17:45--20:00, 20:45--21:00, 22:30--23:00, and 23:15--23:30. In contrast, under the proposed \ac{mbro} framework, price deviations are distributed across multiple bounds, so moderate deviations occur more often while extreme deviations are less frequent, yielding a less conservative and more realistic representation of market prices. For example, in the optimistic strategy, the \ac{mbro} model allows low deviations in 16 periods, medium deviations in 4 periods, and high deviations in only 2 periods (see \Cref{table:Data_Budget}). 
%Illustratively, in periods 0:00--0:15, 16:15--16:45, and 17:45--18:00 the \ac{dam} price exhibits moderate deviations (38.6\%, 33.9\%, 26.9\%, and 25.6\%, respectively) relative to the median. High deviations occur only in 17:15--17:30 and 20:15--20:30, with deviations of 39.8\% and 20.5\%, respectively. 
As the bidding strategy becomes more conservative (balanced and pessimistic), price deviations occur in more periods in both formulations. Nevertheless, a clear difference remains: the single-bound \ac{ro} model tends to push prices to their worst-case values more often, resulting in very conservative trajectories with abrupt jumps, whereas the \ac{mbro} model yields smoother behavior, with frequent low deviations and relatively rare extreme deviations.

\begin{figure}[t!]
    \centering
        {\includegraphics[width=\linewidth]{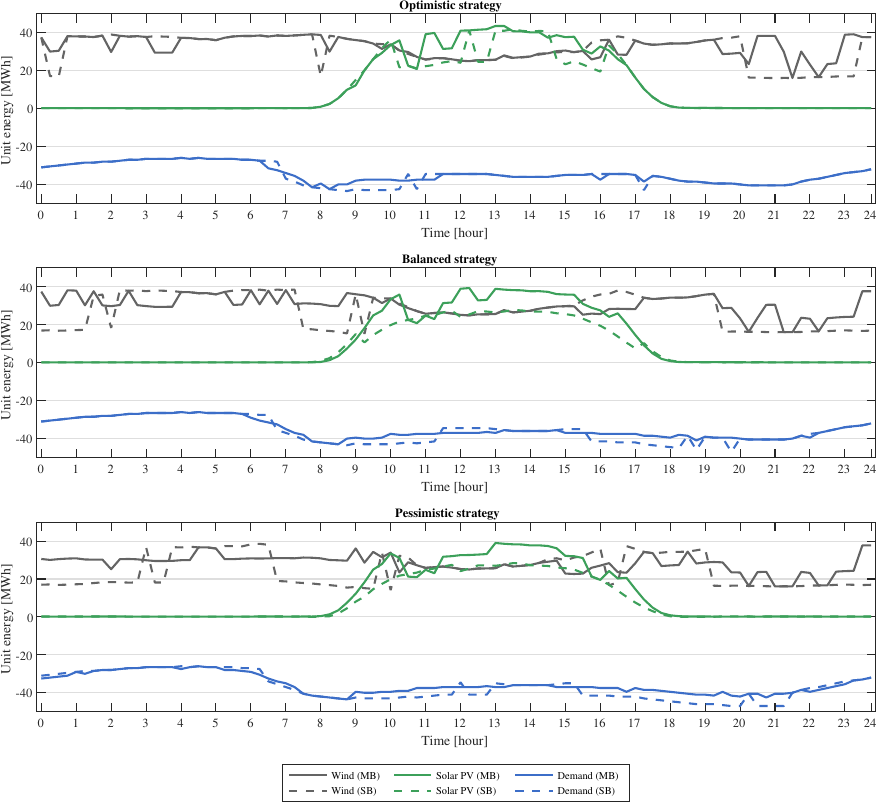}}
    \caption{Case 3 -- Unit energy in multi-bound and single-bound for different uncertainty handling strategies.}
    \label{fig:Unit_Energy_Case3}
    %\vspace{-2em}
\end{figure}

\Cref{fig:Unit_Energy_Case3} compares the energy generation and consumption of the uncertainty-affected \ac{rvpp} units obtained with the proposed \ac{mbro} framework and the classic \ac{ro} model under different uncertainty-handling strategies. In the \ac{ro} model, uncertainty is represented through a single worst-case realization, which translates into only acute decreases in wind farm and solar \ac{pv} production and steep increases in demand consumption. For example, under the optimistic strategy, the wind farm output exhibits negative deviations in periods 0:15--0:45, 8:00--8:15, and 20:15--23:15, with deviations ranging from 55.9\% to 59.1\% relative to the upper forecast bound. In contrast, the proposed \ac{mbro} framework spreads deviations across multiple bounds, so low deviations occur more frequently and very sharp deviations occur less often. 
%This leads to a more realistic representation of uncertainty over the full scheduling horizon and, consequently, less conservative outcomes. For instance, for the wind farm under the optimistic strategy, low deviations occur in 16 periods with deviations between 21.4\% and 25.6\%; moderate deviations occur only in periods 20:15--20:30, 22:00--22:15, and 22:30--22:45 with deviations between 38.6\% and 39.9\%; and sharp deviations occur only in periods 21:30--21:45 and 22:15--22:30 with deviations between 57.9\% and 59.1\%. 
When moving from the optimistic to the balanced and pessimistic strategies, the \ac{ro} model becomes increasingly conservative, driving uncertain generation/consumption close to their worst-case levels in most periods. By comparison, the proposed \ac{mbro} approach remains less conservative, with uncertainty primarily reflected through low and medium deviations and with extreme deviations appearing in relatively few periods.
\vspace{20em}

\begin{figure}[t!]
    \centering
        {\includegraphics[width=\linewidth]{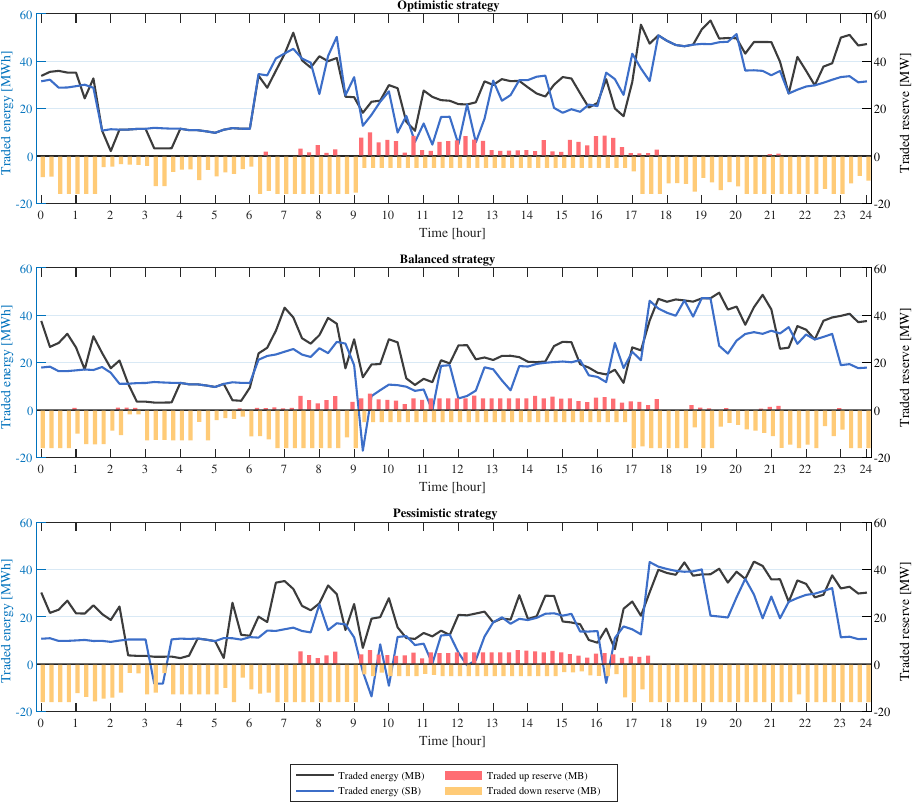}}
    \caption{Case 3 -- Total traded energy and reserve of RVPP in multi-bound and single-bound for different uncertainty handling strategies.}
    \label{fig:Traded_energy_Case3}
    %\vspace{-2em}
\end{figure}

\Cref{fig:Traded_energy_Case3} compares the traded energy of the \ac{rvpp} obtained with the proposed \ac{mbro} framework and the classic \ac{ro} model under different uncertainty-handling strategies. The traded reserve of the \ac{rvpp} under the proposed \ac{mbro} framework is also shown in the figure. The traded reserve of the \ac{rvpp} under the \ac{ro} model, which was previously reported in \Cref{fig:Traded_energy_Case2}, is omitted here to avoid cluttering the figure. Overall, the \ac{ro} approach leads to a much more conservative trading strategy for the \ac{rvpp} in both the \ac{dam} and the \ac{srm} compared to the \ac{mbro} approach, confirming the observations discussed above. 
%For example, under the optimistic strategy, the traded energy of the \ac{rvpp} in the \ac{ro} approach (blue line in \Cref{fig:Traded_energy_Case3}) is lower than the corresponding traded energy under the \ac{mbro} approach (black line in \Cref{fig:Traded_energy_Case3}) in most periods. This conservatism becomes even more pronounced under the balanced and pessimistic strategies in the \ac{ro} approach. For instance, under the pessimistic strategy, between periods 9:00--12:30, the \ac{ro} model accounts for extreme worst-case energy deviations of the \ac{rvpp} units, leading the \ac{rvpp} to trade zero or very low energy and, in some of these periods, to purchase energy in the \ac{dam}. In contrast, although the traded energy of the \ac{rvpp} decreases under the pessimistic strategy relative to the optimistic and balanced strategies, the proposed \ac{mbro} approach provides a more realistic and less conservative solution: under the pessimistic strategy, traded energy is only moderately affected in most of these periods, and severe reductions occur less frequently. A similar behavior is observed for reserve trading. For example, under the pessimistic strategy, comparing the reserve results under the \ac{mbro} approach (\Cref{fig:Traded_energy_Case3}) with those under the \ac{ro} approach (\Cref{fig:Traded_energy_Case2}) shows that both upward and downward reserves in the \ac{ro} model are strongly affected and reduced, whereas under the \ac{mbro} approach the \ac{rvpp} provides higher upward and downward reserves in the \ac{srm}.
\vspace{20em}

\Cref{table:Economic_Case3} compares the economic results of \ac{rvpp} participation in the \ac{dam} and \ac{srm}, obtained using the proposed \ac{mbro} framework and the classic \ac{ro} model under different uncertainty-handling strategies. The table reports the revenue from trading energy and reserve, the operation cost of \ac{rvpp} units, the robust cost (i.e., the third line in the objective function~\eqref{MILP: Obj}), and the resulting profit, defined as revenue minus the operation and robust costs. It also reports the computational time of the simulations. Overall, the \ac{mbro} model returns higher profit and lower robust cost than the \ac{ro} model across all uncertainty-handling strategies. To avoid inflated relative improvements when the \ac{ro} profit is small, we normalize the profit difference by the \ac{ro} optimistic profit (constant denominator across strategies). With this normalization, the \ac{mbro} profit under the optimistic, balanced, and pessimistic strategies is 24.9\%, 42.8\%, and 49.2\% higher than that of the \ac{ro} model, respectively. Moreover, the robust cost of the \ac{mbro} model is lower than that of the \ac{ro} model; under the optimistic, balanced, and pessimistic strategies, the corresponding reductions are 26.7\%, 19.3\%, and 8.5\%, respectively. Finally, within the \ac{ro} model, adopting the more conservative balanced and pessimistic strategies (relative to the optimistic strategy) reduces the \ac{rvpp} profit by 52.9\% and 87.6\%, respectively. In contrast, the corresponding profit reductions in the \ac{mbro} approach are more moderate, at 28.0\% and 50.7\%. Regarding the computational time, the \ac{ro} approach typically solves in less than 10~seconds for all uncertainty-handling strategies. The proposed \ac{mbro} approach requires higher computational effort, but for the considered \ac{rvpp} it is usually solved in less than 1~minute across the different strategies. This comparison indicates that, while the \ac{mbro} approach provides greater flexibility for \ac{rvpp} operators to handle multiple sources of uncertainty, its computational burden remains moderate and does not increase excessively compared to the simpler \ac{ro} approach. Therefore, the proposed \ac{mbro} framework is efficient and suitable for performing multiple analyses before submitting the final \ac{rvpp} bid to the market. These results highlight the effectiveness of the proposed \ac{mbro} approach for \ac{dam} and \ac{srm} participation. In comparison, the \ac{ro} model leads to overly conservative solutions that may cause substantial profit reductions in market participation. 
%By handling multiple sources of uncertainty more effectively, the proposed approach provides a less conservative and more profitable solution.

\begin{table}[t!]
  \centering
    %\small
  \caption{RVPP economic results in multi-bound and single-bound for different uncertainty handling strategies.}
  \begin{threeparttable}
  \begin{tabular}{cclccccc}
    \toprule
    
    \multicolumn{1}{c}{\textbf{Model}} 
     & \multicolumn{1}{c}{\textbf{Strategy}} 
    && \multicolumn{1}{c}{\text{Profit}}
    & \multicolumn{1}{c}{\text{Revenue}} 
    & \multicolumn{1}{c}{\text{Operation cost}} 
    & \multicolumn{1}{c}{\text{Robust cost}} 
    & \multicolumn{1}{c}{\text{Computational time}}

    \\
    \multicolumn{1}{c}{\textbf{}} & {\textbf{}} && \text{[k€]} & \text{[k€]} & \text{[k€]} & \text{[k€]}  & \text{[s]}
    \\
    
    \specialrule{2pt}{2pt}{2pt}     
    \multirow{3}{*}{SB approach} &  {Optimistic} && 19.85    & 46.71  & 20.98  & 5.88 & 3.0   \\ \cmidrule{2-8}
    & {Balanced} && 9.35 & 36.37 & 19.22  & 7.80 & 6.5     \\ \cmidrule{2-8}
    & {Pessimistic} && 2.45  & 28.55   & 17.76  & 8.34 & 4.7 \\ [0.2em]

    \specialrule{1pt}{1pt}{1pt}     
    \multirow{3}{*}{MB approach} & {Optimistic} && 24.81    & 51.01  & 21.89  & 4.31 & 59.5     \\ \cmidrule{2-8}
    & {Balanced} && 17.85 & 44.87   & 20.73 & 6.29 & 38.8     \\ \cmidrule{2-8}
    & {Pessimistic} && 12.22  & 39.78   & 19.93 & 7.63 & 17.9     \\ [0.2em]
    
\bottomrule
  \end{tabular}
\vspace{1mm}
\end{threeparttable}
  \label{table:Economic_Case3}
\end{table}

\section{Conclusion}
\label{sec:Conclusion}

{\color{black}This paper proposes an \ac{mbro} framework for the optimal bidding of \ac{rvpp} in energy and reserve markets, designed to align with the ongoing shift toward higher temporal market resolution (e.g., quarter-hourly scheduling).} By deriving an \ac{milp} reformulation of the problem, the proposed method provides a mechanism to manage multiple uncertainties, including electricity prices, \ac{ndrs} production, and demand, with greater flexibility than classic approaches. Unlike standard \ac{ro}, the \ac{mbro} framework allows for granular control over intra-hourly uncertainty exposure, ensuring protection across a wider range of time intervals without resorting to overly conservative worst-case assumptions. Three main remarks can be drawn from the results of the case studies presented and discussed. First, operational analysis demonstrates that \ac{drs} effectively compensate for renewable fluctuations, while \ac{ess} and hydro units anchor the provision of ancillary services. Under uncertainty, the \ac{rvpp} strategy shifts toward distributing reserve commitments across more intervals with lower individual magnitudes. Second, the comparison of market resolutions highlights the critical importance of granularity; the 15-minute model captures trading dynamics and flow reversals smoothed out by hourly frameworks, with volume discrepancies ranging from 18.0--34.2\% for day-ahead energy, 28.7--65.6\% for upward reserve, and 10.1--16.3\% for downward reserve, based on the scenarios considered. Third, the proposed \ac{mbro} approach significantly outperforms classic \ac{ro} by distinguishing between frequent but moderate deviations and rare, \textit{extreme} anomalies. This strategic flexibility significantly increases profitability across different strategies while maintaining high computational efficiency, with solve times remaining under one minute for all cases considered.

Future work will extend this framework to explicitly model the correlations and non-linear interactions between different uncertainty parameters, further refining the representation of real-world market dynamics.

\appendix
\section*{Appendix: Linear Reformulation of the Protection Functions}
\label{appendix:A}

\setcounter{equation}{0} % Reset equation counter
\renewcommand{\theequation}{A.\arabic{equation}} % Custom format A1, A2,...

To reformulate the protection function of the objective function in~\eqref{Uncertainty: Obj} as a linear optimization problem~\eqref{Protection_Function_obj}, we utilize the following proposition based on~\cite{bertsimas04}.

\textbf{Proposition 1}:
The value of the protection function in the third line of~\eqref{Uncertainty: Obj} is equivalent to the optimal objective value of the linear optimization problem~\eqref{Protection_Function_obj}.

\textbf{Proof}:
The linear maximization problem defined in~\eqref{Protection_Function_obj} maximizes its objective by setting the auxiliary variables corresponding to the largest coefficients to 1, contingent on constraint feasibility. Specifically, the optimal solution assigns exactly $\Gamma^{DA}_k$ of the $z^{DA}_{k,t}$ variables to 1 for each bound $k$, provided that the limitation on the summation of the auxiliary variables $z^{DA}_{k,t}$ across all uncertainty bounds for any given time period in~\eqref{Protection_obj: con4} is satisfied. Analogous logic applies to the reserve variables, where $\Gamma^{SR,\uparrow}_k$ of the ${z^{SR,\uparrow}_{k,t}}$ variables and $\Gamma^{SR,\downarrow}_k$ of the ${z^{SR,\downarrow}_{k,t}}$ variables are set to 1, subject to similar summation limits in constraints~\eqref{Protection_obj: con5} and~\eqref{Protection_obj: con6}, respectively. This mechanism is mathematically equivalent to the selection of the worst-case time sets $\left\{ \mathscr{T}^{DA}_k \middle| \; \left| \mathscr{T}^{DA}_k \right| = \Gamma^{DA}_k \right\}$, $\left\{ \mathscr{T}^{SR,\uparrow}_k \middle| \; \left| \mathscr{T}^{SR,\uparrow}_k \right| = \Gamma^{SR,\uparrow}_k \right\}$, and $\left\{ \mathscr{T}^{SR,\downarrow}_k \middle| \; \left| \mathscr{T}^{SR,\downarrow}_k \right| = \Gamma^{SR,\downarrow}_k \right\}$ utilized in the protection function in~\eqref{Uncertainty: Obj}. \hfill$\blacksquare$
\vspace{3mm}

Additionally, to reformulate the protection function of the \ac{ndrs} constraint in~\eqref{Uncertainty: NDRES} as a linear optimization problem~\eqref{Protection_Function_NDRES}, we utilize the following proposition.

\textbf{Proposition 2}:
The value of the protection function in the right-hand side of~\eqref{Uncertainty: NDRES} is equivalent to the optimal objective value of the linear optimization problem~\eqref{Protection_Function_NDRES}.

\textbf{Proof}:
The linear maximization problem defined in~\eqref{Protection_Function_NDRES} maximizes the deviation for the current period $t^\prime = t$ by setting the auxiliary variable $z_{r,k,(t^\prime=t)}$ to 1, contingent on constraint feasibility. Specifically, the optimal solution assigns $z_{r,k,(t^\prime=t)} = 1$ for a chosen bound $k$ if and only if the remaining uncertainty budget (calculated as $\Gamma_{r,k} - \sum_{t^\prime \neq t} z_{r,k,t^\prime}^{*}$) is sufficient, provided that the limitation on the summation of the auxiliary variable $z_{r,k,(t^\prime=t)}$ across all uncertainty bounds in~\eqref{Protection_NDRES: con2} is satisfied. This mechanism is mathematically equivalent to determining whether period $t$ can be included in the worst-case time set $\mathscr{T}_{r,k}$ for a specific bound $k$ utilized in the protection function in~\eqref{Uncertainty: NDRES}. \hfill $\blacksquare$

%\section*{Acknowledgments}
%The authors wish to thank Comunidad de Madrid for the financial support to PREDFLEX project (TEC-2024/ECO-287), through the R\&D activity programme Tecnologías 2024.

%\clearpage
\begin{footnotesize}
\setlength{\bibsep}{5pt}
\bibliography{refs.bib}
\bibliographystyle{elsarticle-num.bst} 

\end{footnotesize}

\end{document}